\documentclass[11pt,a4paper]{article}
\usepackage{jheppub}
\usepackage{epsfig}
\usepackage{amssymb}
\usepackage{amsmath}
\usepackage{float}
\usepackage{subfigure, amsmath,graphicx}

\newcommand{\bea}{\begin{eqnarray}}
\newcommand{\eea}{\end{eqnarray}}
\newcommand{\bean}{\begin{eqnarray*}}
\newcommand{\eean}{\end{eqnarray*}}
\newcommand{\nn}{\nonumber \\}

\def\W #1{\widetilde{#1}}

\def\a{{\alpha}}

\def\b{{\beta}}

\def\la{\lambda}
\def\eps{\epsilon}

\def\bvev#1{\left[ #1 \right]}

\def\Spbb{\bvev}

\def\Sl{\sum\limits}
\def\Label#1{\label{#1}%
  \smash{\hbox to0pt{\raise1ex\hbox{\tiny[#1]}\hss}}}

\def\Label#1{\label{#1}%
  \smash{\hbox to0pt{\raise1ex\hbox{\tiny[#1]}\hss}}}

\title{An algebraic approach to BCJ  numerators}

\author[a,b]{Chih-Hao Fu,}
\author[c,d]{Yi-Jian Du,}
\author[a,d]{Bo Feng,}

\affiliation[a]{Center of Mathematical Science, Zhejiang
University\\38 Zheda Road Hangzhou, 310027 P.R China}

\affiliation[b]{Department of Electrophysics, National Chiao Tung University\\
1001 University Street, Hsinchu, Taiwan, R.O.C.}

\affiliation[c]{Department of Physics and Center for Field Theory
and Particle
 Physics, Fudan University,\\ Shanghai 200433, P.R China}

\affiliation[d]{Zhejiang Institute of Modern Physics, Zhejiang
University\\38 Zheda Road Hangzhou, 310027 P.R China}

\emailAdd{zhihaofu@nctu.edu.tw} \emailAdd{yjdu@fudan.edu.cn}
\emailAdd{b.feng@cms.zju.edu.cn}

\date{\today}
\abstract{ One important discovery in recent years is that the total
amplitude of gauge theory can be written as BCJ form where kinematic
numerators satisfy Jacobi identity. Although the existence of such
kinematic numerators is no doubt, the simple and explicit construction is still
an important problem. As a small step, in this note we provide an
algebraic approach to construct these kinematic numerators. Under our
Feynman-diagram-like  construction, the Jacobi identity is manifestly satisfied.
The corresponding color ordered amplitudes satisfy off-shell
KK-relation and off-shell BCJ relation similar to the color ordered
scalar theory. Using our construction, the dual DDM form is also
established.

 }


\begin{document}
\maketitle

\section{Introduction}

Recent studies  have revealed that there are many new structures for
scattering amplitudes unforeseen from lagrangian perspective. One of
such examples is  the color-kinematic duality discovered by Bern,
Carrasco and Johansson\cite{Bern:2008qj} (BCJ). In the work  it was
conjectured that color-ordered amplitudes of gauge theories  can be
rearranged into a form where kinematic numerators satisfy the same
Jacobi identities as the color part does (i.e, the part given by
multiplication of structure constants of gauge group according to
corresponding cubic Feynman diagrams). These forms (we will call
BCJ-form)  lead to very nontrivial linear relations among color
ordered amplitudes\footnote{BCJ relations between color-ordered
amplitudes has been proved in string theory in
\cite{BjerrumBohr:2009rd,Stieberger:2009hq, BjerrumBohr:2010zs,
Mafra:2011kj} and in field theory in \cite{Feng:2010my,
Jia:2010nz,Chen:2011jxa}
 using on-shell recursion relations}, thus we can  reduce
 the number of independent amplitudes to $(n-3)!$.
 A further conjecture  of color-kinematic dual form (BCJ-form)
is that if we replace the color part by kinematic part in the
BCJ-form, we will get corresponding gravity amplitudes.  The
double-copy formulation of tree-level gravity amplitudes is
equivalent\footnote{A proof can be found in \cite{Bern:2010yg}. } to
the Kawai-Lewellen-Tye (KLT) relations \cite{KLT, Bern:1998ug}.
However, unitarity suggests that double-copy formulation may be
generalized beyond tree-level and therefore provides an extremely
useful aspect to understand or calculate
 gravity amplitudes at loop-levels. Recent
discussions on loop-level can be found in \cite{Bern:2008pv,
Bern:2009kd, Bern:2010ue, Carrasco:2011mn,Bern:2011rj,
BoucherVeronneau:2011qv, Bern:2012uf, Bern:2012cd, Bern:2012gh,
Oxburgh:2012zr, Yuan:2012rg, Saotome:2012vy, Boels:2012sy}. Because
these important applications to gravity amplitudes, the simple and
explicit construction of kinematic numerators is very important. In
this paper we show that assuming gauge symmetry provides enough
degrees of freedom, it is possible to construct kinematic numerators
as linear combinations of  contributions coming from cubic graphes,
with vertices given by generalization of the algebraic structure
constant given in \cite{Monteiro:2011pc, BjerrumBohr:2012mg}. This
construction makes many algebraic relations  between numerators,
such as Jacobi identity, KK-relation and BCJ relations, manifest.

%

Another interesting consequence of color-kinematic duality is  that
gauge theory amplitudes may have different forms. Two such examples
are the color-ordered decomposition
$\mathcal{A}_{tot}=\sum_{\sigma\in S_{n-1}}
Tr(T^{\sigma_{1}}T^{\sigma_{2}}\dots T^{\sigma_{n}})\, A(\sigma)$
(which we will call "Trace form") and the form discovered by Del
Duca, Dixon and Maltoni (DDM) \cite{DelDuca:1999rs}
$\mathcal{A}_{tot}= \sum_{\sigma\in
S_{n-2}}f^{1\sigma_{2}x_{1}}f^{x_{1}\sigma_{3}x_{2}} \dots
f^{x_{n-3}\sigma_{n-1}n}\, A(1,\sigma,n)$ (which we will call the
"DDM form"). The equivalence of two forms gives another proof of
Kleiss-Kuijf (KK) \cite{Kleiss:1988ne} relations of the
color-ordered amplitudes\footnote{In \cite{DelDuca:1999rs}, the DDM
form was derived using the properties of Lie algebra. However, it
can also be derived \cite{Du:2011js} using KLT formulation of
Yang-Mills amplitude \cite{Bern:1999bx}.}. Within the
color-kinematic duality, it is natural to have the "Dual Trace form"
and "Dual DDM form" as discussed in \cite{Bern:2010yg,Bern:2011ia}.
However, unlike the Trace form and DDM form, the dual form does not
have very simple construction for the dual color part. In this
paper, we will give a partial construction of the dual color part.

This paper is organized as follows. In section \ref{sec:algebra} we
introduce the Lie algebra of general diffeomorphism in Fourier
basis. Upon the sum of  cyclic permutations of  the structure
constant, we get  Yang-Mills $3$-point vertex. Section
\ref{sec:algorithm} is our main part where the construction of
kinematic numerators is given. We start with two examples, the
4-point numerator and 5-point numerator, where explicit calculations
are given. Then we give a general frame for our construction. In
section \ref{sec:identities} we discuss relations, such as KK and
off-shell BCJ relations, among quantities defined in section
\ref{sec:algorithm}. In section \ref{sec:DDM} we derive the dual DDM
form using relations from previous section. A few comments on
relations between different formulations of Yang-Mills amplitudes
are given in section \ref{sec:formulations}. After a short
conclusion,  a proof of KK relation using off-shell recursion
relation is included in the appendix.


\section{Generators and kinematic structure constant}
\label{sec:algebra}

Our starting point is a generalization of the
diffeomorphism Lie algebra introduced by Bjerrum-Bohr, Damgaard,
Monteiro and O'Connell \cite{Monteiro:2011pc, BjerrumBohr:2012mg}.
The generator is defined as
\begin{equation}
T^{k,a}\equiv e^{ik\cdot x}\partial_{a},\label{eq:generator}
\end{equation}
with label $(k,a)$, where $k$ is a $D$-dimensional vector and
$a$, a Lorentz index. The kinematic structure constant can be read out from commutator
\begin{eqnarray}
[T^{k_{1},a},T^{k_{2},b}] & = & (-i)(\delta_{a}{}^{c}k_{1b}-\delta_{b}{}^{c}k_{2a})\,
e^{i(k_{1}+k_{2})\cdot x}\partial_{c} \\
 & = & f^{(k_{1},a),(k_{2},b)}{}_{(k_{1}+k_{2},c)}\, T^{(k_{1}+k_{2},c)}. \nonumber
\end{eqnarray}
In the following we shall use a shorthand notation by writing
 $f^{(k_{1},a),(k_{2},b)}{}_{(k_{1}+k_{2},c)}$
as $f^{1_{a},2_{b}}{}_{(1+2)^{c}}$. The upper and lower scripts
of Lorentz indices $a$, $b$ and $c$ are introduced to distinguish
whether the corresponding generators are contravariant or covariant
under Lorentz symmetry. Jacobi identity coming from  cyclic sum
 of the commutator $[[T^{k_{1},a},T^{k_{2},b}],T^{k_{3},c}]$ is given by
\begin{equation}
f^{1_{a},2_{b}}{}_{(1+2)^{e}}f^{(1+2)_{e},3_{c}}{}_{(1+2+3)^{d}}
+f^{2_{b},3_{c}}{}_{(2+3)^{e}}f^{(2+3)_{e},1_{a}}{}_{(1+2+3)^{d}}
+f^{3_{c},1_{a}}{}_{(1+3)^{e}}f^{(1+3)_{e},2_{b}}{}_{(1+2+3)^{d}}=0.
\end{equation}

\begin{figure}[!h]
  \centering
 \includegraphics[width=8cm]{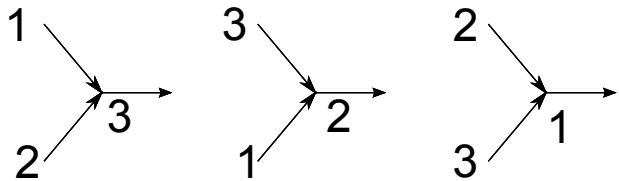}
 \caption{From left to right, three diagrams represent $f^{12}_{~~~3}, f^{31}_{~~~2},
f^{23}_{~~~1}$ in Eq.(\ref{eq:3-pt-vertex}), where we used
arrows to distinguish  upper from lower indices.
}~~~\label{fig-3-vertex}
\end{figure}

To relate structure constants to  Feynman rules, we need to lower or raise
Lorentz indices by contracting with Minkowski
metric. For example
\begin{eqnarray}
f^{1_{a},2_{b}}{}_{(1+2)_{c}} & \equiv & f^{1_{a},2_{b}}{}_{(1+2)^{e}}\,\eta_{ec}=(-i)(\eta_{ac}k_{1b}-\eta_{bc}k_{2a}),~\label{eq:lowered-fabc}
\end{eqnarray}
The index-lowered structure constant (\ref{eq:lowered-fabc}) does
not enjoy cyclic symmetry. However summing over cyclic permutations
of $k_{1}$, $k_{2}$ and $k_{3}=-k_{1}-k_{2}$ produces the familiar color
ordered $3$-point Yang-Mills vertex
\begin{equation}
\frac{1}{\sqrt{2}}\left(f^{1_{a},2_{b}}{}_{-3_{c}}+f^{3_{c},1_{a}}{}_{-2_{b}}
+f^{2_{b},3_{c}}{}_{-1_{a}}\right)=\frac{i}{\sqrt{2}}
\left[\eta_{ab}(k_{1}-k_{2})_{c}+\eta_{bc}(k_{2}-k_{3})_{a}
+\eta_{ca}(k_{3}-k_{1})_{b}\right].\Label{eq:3-pt-vertex}
\end{equation}
Three terms at the left handed side of (\ref{eq:3-pt-vertex})
can be represented by the three arrowed graphs in Figure
\ref{fig-3-vertex}. In this representation, two upper indices $a,b$
of $f^{ab}_{~~~c}$ are denoted by arrows pointing towards the vertex while
lower index $c$ is denoted by an arrow leaving the vertex. These three
terms are related to each other by counter-clockwise cyclic rotation,
thus from the left to right, they represent $f^{12}_{~~~3}, f^{31}_{~~~2},
f^{23}_{~~~1}$.

Note that when expressed in terms of index-lowered structure constants, Jacobi
identity becomes
\begin{equation}
f^{1_{a},2_{b}}{}_{(1+2)_{e}} \eta^{e\W e}f^{(1+2)_{\W
e},3_{c}}{}_{(1+2+3)^{d}} +f^{2_{b},3_{c}}{}_{(2+3)_{e}}\eta^{e\W
e}f^{(2+3)_{\W e},1_{a}}{}_{(1+2+3)^{d}}
+f^{3_{c},1_{a}}{}_{(1+3)_{e}}\eta^{e\W
e}f^{(1+3)_{e},2_{b}}{}_{(1+2+3)^{d}}=0.~~~\label{l_Jac}
\end{equation}
When we interpret relations between numerators as Jacobi
identities, the Minkowski metric $\eta^{e\W e}$ comes from gluon
propagator and connects  two structure constants.
In discussions below we neglect Lorentz indices of structure
constants, which  can be easily recovered from the
context. Contraction of a structure constant $f^{1_{a},2_{b}}{}_{3_{c}}$
with other structure constants should be understood as the same as
contracting a tensor $f^{1,2}{}_{3}$ labelled by legs $1$, $2$,
$3$.


\section{Construction of kinematic numerators}
\label{sec:algorithm}

In this section we present an algorithm to construct the
kinematic numerators that satisfy Jacobi identity as proposed by Bern,
Carrasco and Johansson \cite{Bern:2008qj}. We
demonstrate our method through $4$-point and $5$-point amplitudes,
and then present the general picture for arbitrary
$n$-point amplitudes.

\subsection{kinematic numerators for $4$-point amplitudes}
\label{sec:explicit-cancellation}

For $4$-point amplitudes, we consider two color-ordered ones $A(1234)$ and
$A(1324)$, since rest of amplitudes can be obtained from these
two with the Kleiss-Kuijf (KK) \cite{Kleiss:1988ne} relations. From the
prescription of Bern, Carrasco and Johansson,  $4$-point
color-ordered amplitudes can be divided into contributions of $s$, $t$ and
$u$-channels \cite{Bern:2008qj} \footnote{We follow the sign
convention such that $n_s, n_t, n_u$  correspond to cyclic permutations
of the three external particles with the fourth one  fixed, as they appear in
the Jacobi identity.},
\begin{equation}
A(1234)=\frac{n_{s}}{s}-\frac{n_{u}}{u},\hspace{1cm}A(1324)=-\frac{n_{t}}{t}+\frac{n_{u}}{u}.
\label{eqn:4pt-def-numerator}
\end{equation}
Our goal is to  construct  kinematic (BCJ) numerators
$n_{s}$, $n_{t}$, $n_{u}$ that satisfy Jacobi identity
$n_{s}+n_{t}+n_{u}=0$. Let us first focus on amplitude $A(1234)$.
From color-ordered Feynman rules, amplitude $A(1234)$ contains a
$s$-channel and a $u$-channel graphes with only cubic vertexes. Thus it
 is natural to attribute expressions coming from Feynman rules to numerators $n_{s}$ and
$n_{u}$ respectively. In addition we have a contribution from
color-ordered $4$-point vertex
\begin{equation}
i\eta_{ac}\eta_{bd}-\frac{i}{2}(\eta_{ab}\eta_{cd}+\eta_{ad}\eta_{bc})
=\frac{i}{2}(\eta_{ac}\eta_{bd}-\eta_{ad}\eta_{bc})+\frac{i}{2}
(\eta_{ac}\eta_{bd}-\eta_{ab}\eta_{cd}),~~~\label{4-vertex}
\end{equation}
(where the Lorentz indices of
particles $1,2,3,4$  are $a,b,c,d$,).  We attribute
the first and the second terms of (\ref{4-vertex}) at the right handed side to $n_{s}$ and
$n_{u}$\footnote{The reason for assigning the first term with $\eta_{ad}
\eta_{bc}$ to $s$-channel can be understood as collecting
contributions that carry the same color dependence as $s$-channel graph
from the complete $4$-point Yang-Mills vertex.}. Using propagator ${-i\eta_{\mu\nu}\over p^2}$, the
$s$-channel numerator $n_{s}^{*}$ can be read out
%
\begin{eqnarray}
n_{s}^{*} & = & -\frac{i}{2}\left[f^{1,2}{}_{-e}+(f^{e,1}{}_{-2}
+f^{2,e}{}_{-1})\right]\cdot\left[f^{3,4}{}_{e}+(f^{e,3}{}_{-4}+f^{4,-e}{}_{-3})\right]\\
 &  & +\frac{i}{2}s\,(\eta_{ac}\eta_{bd}-\eta_{ad}\eta_{bc}), \nonumber
\end{eqnarray}
where we used equation (\ref{eq:3-pt-vertex}) to express $3$-point
Yang-Mills vertex in terms of kinematic structure constants. 
A star sign of $n_s^*$ was introduced to denote  quantities
that have not been contracted with polarization vectors.  Expanding product of
kinematic structure constants in the first line yields the following nine terms
\bea
& f^{1,2}{}_{-e}(f^{e,3}{}_{-4}+f^{4,-e}{}_{-3})+(f^{e,1}{}_{-2}+f^{2,e}{}_{-1})f^{3,4}{}_{e}
\label{eq:product-expansion} \\
& +f^{1,2}{}_{-e}f^{3,4}{}_{e}+(f^{e,1}{}_{-2}+f^{2,e}{}_{-1})(f^{e,3}{}_{-4}+f^{4,-e}{}_{-3}),
\nonumber
\eea
\begin{figure}[!h]
  \centering
 \includegraphics[width=15cm]{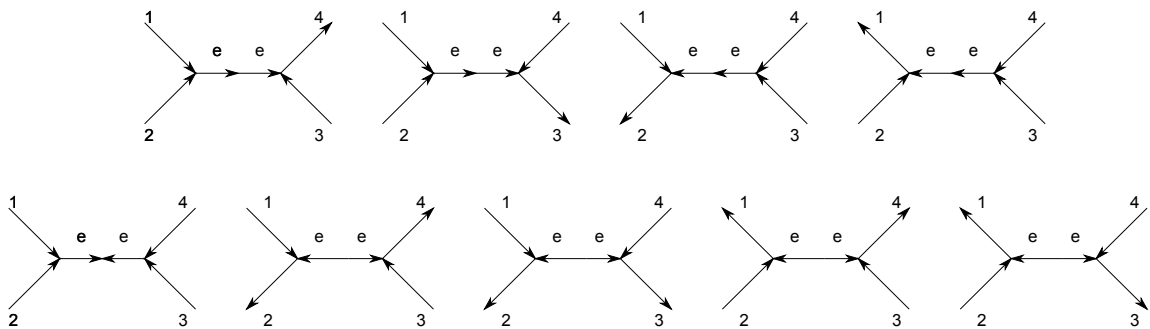}
 \caption{Graphical representation of contributions from
 Eq. (\ref{eq:product-expansion}) where
 the first four graphes correspond to contributions from the first line
of the equation, and the remaining five graphes, from  the second line of the equation.
 }~~~\label{fig-4-ns}
\end{figure}
which can be represented by the graphs in Figure
\ref{fig-4-ns}. From these graphes, several information can be read out.

 First we note that contraction of the repeated index $e$
leads to consistent arrow directions for internal lines
in first four graphs but inconsistent arrow directions for internal line in
the remaining five graphs.
As we will see in discussions below, contributions from
consistent contractions
satisfy the Jacobi identity of kinematic structure
constants $f^{ab}{}_c$ while inconsistent contractions do not.
Because of  this reason we shall split contributions from Feynman
diagrams consisting of only cubic vertices into two groups: The ``good ones''
with consistent contractions and the ``bad ones'' with at least one
inconsistent contractions.

Secondly, we note that a ``good'' graph
can  only have one outgoing arrow among all external particles.
The unique external particle line carrying outgoing arrow
plays an important role when we consider identities among graphs. In particular,
we shall see that Jacobi identity is separately satisfied
among graphs that have same outgoing leg.

Based on  above observations, we can write numerator $n_s^*=G+ X$ where $G$ is
contributions from good graphs and $X$ is contributions from bad graphs and four-point vertex
$\frac{i}{2}s\,(\eta_{ac}\eta_{bd}-\eta_{ad}\eta_{bc})$. As we
will explain in section \ref{sec:elim-cont}, remainder $X$ can be eliminated through averaging
procedure.

However, for the simple 4-point amplitude, we can do better by digging out some good part from
the ``bad contribution''. We note that we are allowed to freely translate between
upper and lower script structure constants through the identities
\begin{eqnarray}
(f^{e,1}{}_{-2}+f^{2,e}{}_{-1})-f^{1,2}{}_{-e} & = & -i
\eta_{ab}(-k_{1}+k_{2})_{e}
+\mathcal{O}(k_{1a},k_{2b}),\label{eq:identity-ab}\\
(f^{e,3}{}_{-4}+f^{4,-e}{}_{-3})-f^{3,4}{}_{e} & = &
-i\eta_{cd}(-k_{3}+k_{4})_{e}
+\mathcal{O}(k_{3c},k_{4d}),\label{eq:identity-cd}
\end{eqnarray}
where $\mathcal{O}_e(k_{1a},k_{2b})$ denotes longitudinal term
$\mathcal{O}_e(k_{1a},k_{2b}) =-i(\eta_{ea}k_{2b}-\eta_{eb}k_{1a})$,
and similarly does $\mathcal{O}_e(k_{3c},k_{4d})$. Both longitudinal
terms do not contribute when they are contracted with physical polarization
vectors of external legs. Multiplying (\ref{eq:identity-ab})
with (\ref{eq:identity-cd}) we obtain the identity
\bea
& f^{1,2}{}_{-e}f^{3,4}{}_{e}+(f^{e,1}{}_{-2}+f^{2,e}{}_{-1})(f^{e,3}{}_{-4}+f^{4,-e}{}_{-3})~~
\label{4-point-try} \\
& \hspace{3cm}
=f^{1,2}{}_{-e}(f^{e,3}{}_{-4}+f^{4,-e}{}_{-3})+(f^{e,1}{}_{-2}
+f^{2,e}{}_{-1})f^{3,4}{}_{e}-(t-u)\eta_{ab}\eta_{cd} \nn &
+\mathcal{O}(k_{1a},k_{2b},k_{3c},k_{4d}), \hspace{3cm} \nonumber
\eea
where
$\mathcal{O}(k_{1a},k_{2b},k_{3c},k_{4d})=\mathcal{O}_e(k_{3c},k_{4d})\cdot
\mathcal{O}_e(k_{1a},k_{2b})-i\eta_{ab}\mathcal{O}_e(k_{3c},k_{4d})\cdot
(-k_{1}+k_{2})_{e}-i\eta_{cd}\mathcal{O}_e(k_{1a},k_{2b})\cdot
(-k_{3}+k_{4})_{e}$. Thus  $n_{s}^{*}$ is  given by
\bea
n_{s}^*=& -i\,\left[f^{1,2}{}_{-e}(f^{e,3}{}_{-4}+f^{4,-e}{}_{-3})+(f^{e,1}{}_{-2}+f^{2,e}{}_{-1})f^{3,4}{}_{e}\right] \label{eq:numerator-ns} \\
& + \frac{i}{2} \left[
s\,(\eta_{ac}\eta_{bd}-\eta_{ad}\eta_{bc})+(t-u)\eta_{ab}\eta_{cd}\right]
+\mathcal{O}(k_{1a},k_{2b},k_{3c},k_{4d}),\label{eqn:4pt-ns-star} \nonumber
\eea
and  $n_{u}^*, n_{t}^*$ can be derived from it
by  permutations of
indices $(123)\rightarrow(312)$ and $(123)\rightarrow(231)$
respectively.  To obtain the numerators $n_s, n_u, n_t$
in equation (\ref{eqn:4pt-def-numerator}), we just
need to contract $n_{s}^{*}, n_{u}^{*}, n_{t}^{*}$ with physical
polarization vectors, thus the longitudinal
terms $\mathcal{O}(k_{1a},k_{2b},k_{3c},k_{4d})$  drop out.

Having obtained expressions (\ref{eq:numerator-ns}) we want to check the Jacabi idenity
$n_s+n_t+n_u=0$. First we notice that after contraction,
 contributions from  $\frac{i}{2} \left[
s\,(\eta_{ac}\eta_{bd}-\eta_{ad}\eta_{bc})+(t-u)\eta_{ab}\eta_{cd}
\right]$ will be trivially zero under
cyclic sum. To see contributions from the first line of equation
(\ref{eq:numerator-ns})  give zero,
let us expand the first line of  $n_s^*$ into
\begin{equation}
f^{1,2}{}_{-e}f^{-e,3}{}_{-4}+f^{1,2}{}_{-e}f^{4,-e}{}_{-3}
+f^{e,1}{}_{-2}f^{3,4}{}_{e}+f^{2,e}{}_{-1}f^{3,4}{}_{e}.\label{eq:ns-structure}
\end{equation}
and write down corresponding terms of  $n_u^{*}$ by
permutation $(123)\to (231)$
\begin{equation}
f^{2,3}{}_{-e}f^{-e,1}{}_{-4}+f^{2,3}{}_{-e}f^{4,-e}{}_{-1}
+f^{e,2}{}_{-3}f^{2,4}{}_{e}+f^{3,e}{}_{-2}f^{2,4}{}_{e},\label{eq:nu-structure}
\end{equation}
and similarly terms of $n_t^{*}$ by permutation $(123)\rightarrow(312)$
\begin{equation}
f^{3,1}{}_{-e}f^{-e,2}{}_{-4}+f^{3,1}{}_{-e}f^{4,-e}{}_{-2}
+f^{e,3}{}_{-1}f^{2,4}{}_{e}+f^{1,e}{}_{-3}f^{2,4}{}_{e}.\label{eq:nt-structure}
\end{equation}
When summing these three contributions (\ref{eq:ns-structure}),
(\ref{eq:nu-structure}) and (\ref{eq:nt-structure}) together, the
first terms from each contribution add up to  zero because of the
Jacobi identity derived from cyclic permutations of legs $(123)$,
\begin{equation}
f^{1,2}{}_{-e}f^{-e,3}{}_{-4}+f^{2,3}{}_{-e}f^{-e,1}{}_{-4}+f^{3,1}{}_{-e}
f^{-e,2}{}_{-4}=0.\label{eq:4pt-Jacobi1}
\end{equation}
Adding up the rest three terms from each contribution again gives zero by
following three Jacobi identities (\ref{eq:4pt-Jacobi2}),
(\ref{eq:4pt-Jacobi3}) and (\ref{eq:4pt-Jacobi4}),
\bea
f^{1,2}{}_{-e}f^{4,-e}{}_{-3}+f^{2,4}{}_{-e}f^{1,-e}{}_{-3}+f^{4,1}{}_{-e}f^{2,-e}{}_{-3}
& = & 0,\label{eq:4pt-Jacobi2}
\\
f^{3,1}{}_{-e}f^{4,-e}{}_{-2}+f^{1,4}{}_{-e}f^{3,-e}{}_{-2}+f^{4,3}{}_{-e}f^{1,-e}{}_{-2}
& = & 0,\label{eq:4pt-Jacobi3}
\\
f^{2,3}{}_{-e}f^{4,-e}{}_{-1}+f^{3,4}{}_{-e}f^{2,-e}{}_{-1}+f^{4,2}{}_{-e}f^{3,-e}{}_{-1}
& = & 0.\label{eq:4pt-Jacobi4} \eea
which correspond to fixed leg $3,2,1$.

It is easy to see that when expressed graphically,  terms in
equation (\ref{eq:4pt-Jacobi1}) shall all have
the outgoing arrow on leg $4$,
and similarly terms in equations (\ref{eq:4pt-Jacobi2}),
(\ref{eq:4pt-Jacobi3}) and
(\ref{eq:4pt-Jacobi4}) shall all have the outgoing arrows on
 legs $3$, $2$ and leg $1$ respectively.
Thus  Jacobi identity can be
translated as the  sum of three
 graphs related to each other by
 cyclic permutations with a fixed leg having
 outgoing arrow.


\subsubsection{Eliminating contact terms}
\label{sec:elim-cont}
In the discussion above we demonstrated explicitly that
contributions from cubic and quartic diagrams together
give rise to numerators that satisfy the Jacobi identity.
While "good parts" of these contributions satisfy the identity
manifestly, the "bad parts", do not. For $4$-point amplitudes since
structure of  amplitudes is simple, we were able to
rewrite these "bad parts" into nicer forms. However
this rewriting becomes rather difficult for higher point amplitudes,
 therefore we resort
to an alternative way to solve the problem.
The idea is the following. Since the numerator such as
$n_{s}$ is calculated from contracting $n_{s}^{*}$
with polarization vectors, in a gauge theory we have the freedom
to choose different gauges (i.e., different polarization vectors).
Using this freedom we can eliminate "bad contributions" and keep only
"good contributions", thus the final result will satisfy Jacobi identity
manifestly.

Now we demonstrate the idea using $4$-point amplitudes.
For simplicity let us abuse the notation a bit by writing
\bea n_s(q)=\eps_1^{a_1}(q_1)...\eps_n^{a_n}(q_n) n_{s}^*\equiv
\eps(q)\cdot n_s^* ~, \eea
where  $q$ represents the set of reference momenta $\{ q_{1}, q_{2}, \dots , q_{n} \}$
collectively.
Using the notation that the good contribution given by equation
(\ref{eq:ns-structure}), (\ref{eq:nu-structure}) and (\ref{eq:nt-structure})
as $\tilde{n}_{s}^{*}$,
$\tilde{n}_{u}^{*}$ and $\tilde{n}_{t}^{*}$, we have
\bea n_{s}(q)& =& -i\eps(q)\cdot \tilde{n}_{s}^{*}
+{i\over 2} s X_{s}(q),
\nonumber \\ X_{s}(q) & \equiv &
\eps(q)\cdot X_s^*= \eps(q)\left\{
(\eta_{ac}\eta_{bd}-\eta_{ad}\eta_{bc})+{(t-u)\over
s}\eta_{ab}\eta_{cd} \right\}\eea
and similarly for $n_{u}(q), n_t(q), X_{u}(q)$, and $X_{t}(q)$.
Thus the amplitudes are
given by
\bea A(1234)&= &\eps(q)\cdot \left( { -i \tilde{n}_{s}^{*}\over s}-{
-i \tilde{n}_{u}^{*}\over u}+ {i\over 2}X_1^*\right), ~~~~X_1= X_s^*-X_u^*\nonumber \\
A(1324) & = & \eps(q)\cdot \left( -{-i \tilde{n}_{t}^{*}\over t}+{-i
\tilde{n}_{u}^{*}\over u}+ {i\over 2}X_2^*\right), ~~~~X_2=- X_t^*+X_u^*~.\eea
In above expressions,   good contributions into $\tilde{n}^{*}_{i}$,
which satisfy Jacobi identity, have been separated from the
bad contributions  $X^{*}_{i}$ manifestly.
Having done the  reorganization, next step is to eliminate the
 $X^{*}_{i}$ parts. To realize it,  we consider the average of above
two color-ordered amplitudes over three different choices of gauges.
Since $A(1234)$ is invariant under gauge choices,
we can get rid of all $X^{*}_{i}$ parts simultaneously if we impose
 following three conditions
\bea T_3 : \left\{ \begin{array}{l} 1  = c_1+c_2+c_3 \\ 0=
\sum_{i=1}^3 c_i \eps(q_i)\cdot X_1^* \\ 0= \sum_{i=1}^3 c_i
\eps(q_i)\cdot X_2^*
\end{array}\right.~~~\Label{4p-T3}\eea
By gauge invariance, the first condition guarantee that
\bea A(1,2,3,4)= { n_s\over s}-{n_u\over u},~~~ A(1,3,2,4)  =
{-n_t\over t}+{n_u\over u}\eea
where $n_s=\sum_{i=1}^{3} -i c_i  \epsilon (q_i) \cdot\tilde{n}_{s}^{*} $
and similarly for $n_u, n_t$.
Since each $\epsilon (q_i) \cdot\tilde{n}_{s,u,t}$ satisfies Jacobi identity,
so do $n_s, n_u, n_t$. To see that there is indeed a solution for $c_i$, we simply need
to show that the following matrix has nonzero determinant
\bea \left(\begin{array}{ccc} 1 & 1 & 1  \\
\eps(q_1)\cdot X_1^* & \eps(q_2)\cdot X_1^*&\eps(q_3)\cdot X_1^*\\
\eps(q_1)\cdot X_2^* & \eps(q_2)\cdot X_2^*&\eps(q_3)\cdot
X_2^*\end{array}\right)~.\eea
This can be checked by explicit calculations.


\subsubsection{KK vs BCJ-independent basis}
\label{sec:compare-basis}

In  previous section we have showed how to
derive kinematic numerators $n_s, n_t$ and $n_u$ satisfying Jacobi
identity by eliminating bad contributions. In the derivation we considered
the analytic structures of two color ordered amplitudes $A(1234)$ and
$A(1324)$, which serve as a basis when KK-relations \cite{Kleiss:1988ne}
are taken into account. Since there were
two remainders (i.e., the bad contribution part $X$) we need to introduce
 three $c_i$ to achieve our goal. But could we
do better by introducing fewer variables $c_i$?

Let us consider only $A(1234)$. To eliminate its remainder term, we
only need to average over two different gauge choices. The constraint
conditions for $c_i$ are
\bea T_2 : \left\{ \begin{array}{l} 1  = \W c_1+\W c_2 \\ 0=
\sum_{i=1}^2\W c_i \eps(q_i)\cdot X_1^*
\end{array}\right.~~~\Label{4p-T2}\eea
which have the solution $\W c_1={-\eps(q_2)\cdot X_1^*\over \eps(q_1)\cdot
X_1^*-\eps(q_2)\cdot X_1^*}$ and $\W c_2={\eps(q_1)\cdot X_1^*\over
\eps(q_1)\cdot X_1^*-\eps(q_2)\cdot X_1^*}$. Substituting them back, we have
\bea A(1234)= {n_s\over s}-{n_u\over u},~~~ n_s= -i
(\W c_1 \epsilon(q_1) + \W c_2 \epsilon(q_2) ) \cdot \tilde{n}_{s}^{*},
~~~~~n_u= -i
(\W c_1 \epsilon(q_1) + \W c_2 \epsilon(q_2) ) \cdot \tilde{n}_{u}^{*}\eea
Having obtained these two numerators, we can {\it define} an amplitude using
\bea \W A(1324)= -{  -(n_s+n_u)\over t}+ {n_u\over u}\eea
It is easy to check that the amplitude just defined satisfies
fundamental BCJ relation \cite{Bern:2008qj}
by construction,
\begin{eqnarray}
s_{21}A(1234)+(s_{21}+s_{23})\W A(1324) & = & 0,
\end{eqnarray}
Since the same relation is satisfied between physical amplitudes,
$s_{21}A(1234)+(s_{21}+s_{23}) A(1324)=0$, we conclude that $\W
A(1324)=A(1324)$ and in particular, $n_t=-(n_s+n_u)$, i.e., the
kinematic-dual
Jacobi identity we would like to have.

Above discussions show that, because of the BCJ relation for
color-ordered amplitudes, we can use fewer $c_i$ to eliminate
remainders. After doing so, $X_{i}$ in rest of the
color-ordered amplitudes  automatically disappear, i.e.,
\bea \W c_1 \eps(q_1)\cdot X_2^*+\W c_2\eps(q_2)\cdot X_2^*
=0~.~~~\Label{4p-T2-2}\eea
%

Now we have  developed two methods to eliminate remainders
through averaging over KK or BCJ basis of amplitudes.
We need to clarify  the  relation between these two methods.
To do so, let us assume that we have
solution $(c_1, c_2, c_3)$ with gauge choice $\eps(q_3)=\a
\eps(q_1)+\b \eps(q_2)$. This gauge choice can be achieved if
reference spinors of polarization vectors of three particles, for example,
$2,3,4$ are same for gauge choices $\eps(q_1),
\eps(q_2), \eps(q_3)$, but reference spinors of polarization vector of
particle $1$ satisfy the relation $\eps(q_3)=\a
\eps(q_1)+\b \eps(q_2)$. Putting it back to the second equation of
$T_3$ given in (\ref{4p-T3}) and comparing with the second equation
of $T_2$ given in (\ref{4p-T2}), we can write down the following
solution for $T_2$,
\bea \W c_1= c_1+\a c_3+y \eps(q_2)\cdot X_1^*,~~~~ \W c_2= c_2+\b
c_3-y \eps(q_1)\cdot X_1^*\eea
where $y$ is determined by  $\W c_1+\W c_2=1$ to be
$y={(\a+\b-1)c_3\over (\eps(q_1)-\eps(q_2))\cdot X_1^*}$. It is easy
to check that the above indeed constitutes a solution if we assume
 $\a+\b-1=0$. In this case
(\ref{4p-T2-2}) is automatically satisfied
because of the third equation in (\ref{4p-T3}). To see that indeed
$\a+\b=1$, notice that the reference spinors of particle $1$  have relation $\W \mu_{3}=
a\W\mu_{1}+b\W\mu_2$, so
\bean \eps(\W\mu_3)={\la_1\W\mu_3\over \Spbb{1|\W\mu_3}}=\left(
{a\Spbb{1|\W\mu_1}\over a\Spbb{1|\W\mu_1}+b\Spbb{1|\W\mu_2}}\right)
{\la_1\W\mu_1\over \Spbb{1|\W\mu_1}}+\left({b\Spbb{1|\W\mu_2}\over
a\Spbb{1|\W\mu_1}+b\Spbb{1|\W\mu_2}}\right) {\la_1\W\mu_2\over
\Spbb{1|\W\mu_2}}\Longrightarrow \a+\b=1\eean
This explanation shows that solutions $(\W c_1,\W c_2)$ can be taken
as a special case of  solutions $(c_1, c_2, c_3)$.

\subsection{$5$-point numerators}
\label{sec:5pt-example}

For $5$-point amplitudes KK relations reduce the number of
independent color-ordered amplitudes to six. It was shown by Bern,
Carrasco and Johansson \cite{Bern:2008qj} that  these
six amplitudes can be written into following forms with fifteen numerators
suggested by
 possible cubic graphs:
\bea A(12345) & = &
\frac{n_{1}}{s_{12}s_{45}}+\frac{n_{2}}{s_{23}s_{51}}+\frac{n_{3}}{s_{34}s_{12}}
+\frac{n_{4}}{s_{45}s_{23}}+\frac{n_{5}}{s_{51}s_{34}},
\Label{eq:bern-5pt-numerators} \\
A(14325) & = &
\frac{n_{6}}{s_{14}s_{25}}+\frac{n_{5}}{s_{43}s_{51}}+\frac{n_{7}}{s_{32}s_{14}}
+\frac{n_{8}}{s_{25}s_{43}}+\frac{n_{2}}{s_{51}s_{32}},
\nn A(13425) & = &
\frac{n_{9}}{s_{13}s_{45}}+\frac{n_{5}}{s_{34}s_{51}}+\frac{n_{10}}{s_{42}s_{13}}
-\frac{n_{8}}{s_{25}s_{34}}+\frac{n_{11}}{s_{51}s_{42}},\nn
A(12435) & = &
\frac{n_{12}}{s_{12}s_{35}}+\frac{n_{11}}{s_{24}s_{51}}
-\frac{n_{3}}{s_{43}s_{12}}+\frac{n_{13}}{s_{35}s_{24}}
-\frac{n_{5}}{s_{51}s_{43}},\nn
A(14235) & =
&\frac{n_{14}}{s_{14}s_{35}}-\frac{n_{11}}{s_{42}s_{51}}
-\frac{n_{7}}{s_{23}s_{14}}-\frac{n_{13}}{s_{35}s_{42}}-\frac{n_{2}}{s_{51}s_{23}},\nn
A(13245) & = &
\frac{n_{15}}{s_{13}s_{45}}-\frac{n_{2}}{s_{32}s_{51}}
-\frac{n_{10}}{s_{24}s_{13}}-\frac{n_{4}}{s_{45}s_{32}}-\frac{n_{11}}{s_{51}s_{24}},
\nonumber \eea
\begin{figure}[!h]
  \centering
 \includegraphics[width=4cm]{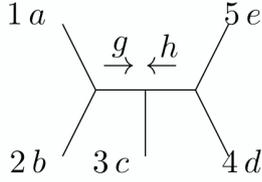}
 \caption{A Feynman diagram contributing to $n_{1}^*$}~~~\label{fig-n1}
\end{figure}

\begin{figure}[!h]
  \centering
 \includegraphics[width=14cm]{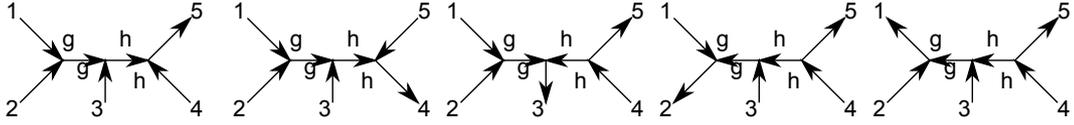}
 \caption{Five  terms with consistent arrow directions contributing to  $n_{1}^*$}~~~\label{fig-5-n1}
\end{figure}

To find  expressions for these $n_i$, as in
the $4$-point amplitudes, we divide contributions from
Feynman rules to good contributions plus a remainder (bad contributions),
\bea A^*(12345) & = &
\frac{n^*_{1}}{s_{12}s_{45}}+\frac{n^*_{2}}{s_{23}s_{51}}+\frac{n^*_{3}}{s_{34}s_{12}}
+\frac{n^*_{4}}{s_{45}s_{23}}+\frac{n^*_{5}}{s_{51}s_{34}}
+X^*(12345)~. \Label{A12345*} \eea
As before we use $*$ to denote quantities that have not been
contracted with  polarization vectors.
The definition of $n_i^*$ and
$X^*$ is the following. First we include all contributions that
contain at least one $4$-point vertex in  Feynman diagrams to $X^*$. For  remaining
Feynman diagrams having only cubic vertices like
Figure \ref{fig-n1} for example, 
we use (\ref{eq:3-pt-vertex})
to translate $3$-point vertices to kinematic structure constants, thus obtain
\footnote{For simplicity we neglect the overall factor ${(-i)^2\over
(\sqrt{2})^3}$, where
$1/\sqrt{2}$ comes from (\ref{eq:3-pt-vertex}) and $(-i)^2$ come from
two propagators.}
\begin{equation}
\left[(f^{g,1}{}_{-2}+f^{2,g}{}_{-1})+f^{1,2}{}_{-g}\right]
\times(f^{-g,3}{}_{h}+f^{-h,-g}{}_{-3}+f^{3,-h}{}_{g})
\times\left[(f^{h,4}{}_{-5}+f^{5,h}{}_{-4})+f^{4,5}{}_{-h}\right]~.
\label{eq:5pt-n1-3pts}
\end{equation}
Expanding (\ref{eq:5pt-n1-3pts}) produces $27$ terms, five terms among them
have consistent arrows in the internal lines (see Figure \ref{fig-5-n1}) (so they
are good contributions).
We  assign these five  terms to $n_1^*$ and the rest
 to $X^*$ (these bad contributions), thus we have
\bea  n_1^*=
f^{1,2}{}_{-g}f^{g,3}{}_{h}(f^{h,4}{}_{5}+f^{5,h}{}_{4})
+f^{1,2}{}_{-g}f^{-h,-g}{}_{-3}f^{4,5}{}_{-h}+(f^{g,1}{}_{-2}
+f^{2,g}{}_{-1})f^{3,-h}{}_{g}f^{4,5}{}_{-h}.\label{eq:5pt-series1}
\eea
It is worth to notice that  five terms in $n_1^*$ correspond to  five
possible assignments of single outgoing arrow to external legs
in graphical representations.
If we use $n_{1,k}^*$ to denote the consistent graph having leg $k$ with
outgoing arrow, for example
$n_{1,3}^*=f^{1,2}{}_{-g}f^{-h,-g}{}_{-3}f^{4,5}{}_{-h}$,
the numerator can be written asa $n_1^*=\sum_{k=1}^5 n_{1,k}^*$.
It is straightforward to see that
numerators from rest of  channels can be written into similar structures.
In particular,  $(-n_{15}^*)$ is found to be the same as
permutation
$(123)\rightarrow(312)$ of $n_{1}^*$
\begin{equation}
-n_{15}^*
=f^{3,1}{}_{-g}f^{g,2}{}_{h}(f^{h,4}{}_{5}+f^{5,h}{}_{4})
+f^{3,1}{}_{-g}f^{-h,-g}{}_{-2}f^{4,5}{}_{-h}+(f^{g,3}{}_{-1}
+f^{1,g}{}_{-3})f^{2,-h}{}_{g}f^{4,5}{}_{-h},\label{eq:5pt-series2}
\end{equation}
and $(-n_4^*)$, the same as permutation $(123)\rightarrow(231)$ of $n_1^*$,
\begin{equation}
-n_4^*=f^{2,3}{}_{-g}f^{g,1}{}_{h}(f^{h,4}{}_{5}+f^{5,h}{}_{4})
+f^{2,3}{}_{-g}f^{-h,-g}{}_{-1}f^{4,5}{}_{-h}+(f^{g,2}{}_{-3}
+f^{3,g}{}_{-2})f^{1,-h}{}_{g}f^{4,5}{}_{-h}.\label{eq:5pt-series3}
\end{equation}
When adding up (\ref{eq:5pt-series1}), (\ref{eq:5pt-series2}) and
(\ref{eq:5pt-series3}), terms with same outgoing leg will add to zero
by Jacobi identity. Thus by our construction, we have
Jacobi identity $n_1^*-n_{15}^*-n_4^*=0$. Similar
argument shows when we permute  $(345)\to (534)$ we will
produce $(-n_3^*)$ and when we permute $(345)\to (453)$ we will
produce $(-n_{12}^*)$. Thus $n_1^*-n_3^*-n_{12}^*=0$ is guaranteed by
 Jacobi identity of kinematic structure constants $f^{ab}{}_{c}$.

%

\subsubsection{Eliminating contact terms}

Having established the form (\ref{A12345*}) as well as similar
expressions for other five amplitudes given in
(\ref{eq:bern-5pt-numerators}), we  construct the $n_i$
given in (\ref{eq:bern-5pt-numerators}) by averaging over different
choices of gauges. Just like for the 4-point amplitudes, we consider
seven gauge choices denoted by $\eps(q_i)$ with $i=1,...,7$ for
polarization vectors under gauge choice
$q_i=\{q_{1,i}, q_{2,i}, q_{3,i}, q_{4,i}, q_{5,i}\}$ and impose
following seven equations for coefficients $c_i$, $i=1,...,7$:
\bea \sum_{i=1}^7 c_i=1,~~~~\sum_{i=1}^7 c_i \eps(q_i)\cdot
X_j=0,~~~j=1,...,6\eea
where  six remainders $X_j$ are those given in (\ref{A12345*}). After solving
$c_i$ from above equations, we can get $n_i$ defined in
(\ref{eq:bern-5pt-numerators}) as following
\bea n_i=\sum_{j=1}^7 c_j \eps(q_j)\cdot n_i^*~.\eea
Since by our construction, $n_i^*$ satisfy Jacobi identity even
before contracting with polarization vectors and $c_j$ are same for all
fifteen $n_i^*$, $n_i$ will too satisfy
Jacobi identity.

In the prescription above, we use the  KK-basis (i.e., the basis
under KK-relation)
and BCJ
relations between amplitudes follow as a consequence of
the Jacobi identities among $n_i$. However, if our focus is
the construction of
these $n_i$ numerators, we can take another logic starting point
using only BCJ-basis (i.e., the basis under BCJ relation). For $5$-point
amplitudes,
 we can take $A(12345)$ and $A(13245)$ as BCJ-basis and
consider averaging over three different gauge choices
\bea c_1+c_2+ c_3=1,~~~~\sum_{i=1}^3 c_i \eps(q_i)\cdot
X^*(12345)=0,~~~\sum_{i=1}^3 c_i \eps(q_i)\cdot X^*(13245)=0\eea
By imposing these conditions we obtain
\bea n_i=\sum_{i=1}^3 c_i \eps_i\cdot
n_i^*,~~~i=1,2,3,4,5,10,11,15~.\eea
Then we construct the remaining seven coefficients using Jacobi
identities
\bea n_6 & = & n_{10}+n_1-n_3-n_4+n_5,~~~~n_7 =
n_2-n_4,~~~~n_8=-n_3+n_5,\nn
n_9 & = &
n_3-n_5+n_6,~~~n_{12}=n_1-n_3,~~~n_{13}=n_1+n_2-n_3-n_4-n_6,~~~n_{14}=-n_2+n_4+n_6\eea
The relation between these two eliminating methods can be understood
similarly to the 4-point example.

\subsection{ $n$-point numerators}

Having  above two examples, it is
straightforward to see
the structure of kinematic numerators for $n$-point amplitudes.
Generically a color-ordered amplitude can be written as
\bea A^*= \sum_i {n_i^*\over D_i} + X^*~~~\Label{n-sep}\eea
where the sum is taken over all cubic graphs. In this expression, $n_i^*$
contain only contributions from cubic  graphs that have
consistent arrow directions. All other contributions
 from cubic  graphs with
inconsistent arrow directions as well as
graphs with at least one $4$-point vertex are assigned to
$X^*$ part. Furthermore, according to which external particle
 has been assigned with the
outgoing arrow in  graphical representation, we can
divide kinematic numerator into
\bea n_i^*= \sum_{k=1}^n n_{i,k}^* , \eea
so that each $n_{i,k}^*$  is represented by a single graph. All  these
$n_{i,k}*$   will have
Jacobi identities among themselves   with different $i$ but same fixed $k$.

Having expressions as   in (\ref{n-sep}), we average
over amplitudes to eliminate the remainder terms $X^*$. This can be done
 through averaging over either KK-basis or BCJ basis.
The average coefficients $c_{i}$ are determined by
 $N_1=(n-2)!+1$ equations
\bea T_{KK}:\left\{ \begin{array}{l} 1  = \sum_{i=1}^{N_1} c_i \\ 0=
\sum_{i=1}^{N_1} c_i \eps(q_i)\cdot X_j^*,~~~~j=1,...,(n-2)!
\end{array}\right.~~~\Label{KK-equ}\eea
for KK-basis or  $N_2=(n-3)!+1$ equations
\bea T_{BCJ}:\left\{ \begin{array}{l} 1  = \sum_{i=1}^{N_2} \W c_i \\
0= \sum_{i=1}^{N_2}\W c_i \eps(q_i)\cdot X_j^*,~~~~j=1,...,(n-3)!
\end{array}\right.~~~\Label{BCJ-equ}\eea
for BCJ-basis. After the averaging we have
$n_j=\sum_{i=1}^{N} c_i\eps(q_i)\cdot n_j^*$.  Other $n_i$'s
which do not show up in the KK-basis or BCJ-basis can be constructed from
various relations including Jacobi identities. From either method we can construct
the  numerators  proposed  by Bern, Carrasco and Johansson
in \cite{Bern:2008qj}.

A technical issue concerning the above averaging procedure is the existence of
solution for equation (\ref{KK-equ}) and (\ref{BCJ-equ}). The existence for lower point
amplitudes can be checked by explicit calculations, but we have not find a proof
for general $n$. In this paper, we will assume their existence.

\section{Fundamental BCJ relations }
\label{sec:identities}

In previous section, we have shown how to construct the kinematic numerator
satisfying the Jacobi identity by averaging over different gauge choices.
 An important step is to separate contributions from Feynman
diagrams to two parts
\bea A^*= \sum_i {\sum_{k=1}^n n^*_{i,k}\over D_i} +
X^* ,~~~\Label{n-sep-new}\eea
where each $n_{i,k}^*$ can be represented by a single consistent arrow
graph with only cubic vertices. Effectively,
we can treat these graphs as if they were built from the
Feynman rules with only  cubic vertices, where the coupling
is given by kinematic structure constant  $f^{ab}{}_{c}$. 
From this point of view we can define
an $n$-point color-ordered   amplitude for given $k$ as
\bea A_{n;k}^* = \sum_i { n^*_{i,k}\over D_i}.~~~\Label{A-k}\eea
The physical amplitude is given by linear combination
of these fixed-$k$ amplitudes
\bea A=\sum_{i=1}^N c_i\eps(q_i)\cdot \sum_{k=1}^n
A_{n;k}^* .~~~~\Label{A-by-Ak}\eea
An important feature of formula (\ref{A-by-Ak}) is that the part
$\sum_{i=1}^N c_i\eps(q_i)$ coming from averaging procedure does not
depend on the color ordering of external particles.

The amplitudes defined in (\ref{A-k}) contain similar algebraic
structure as  these amplitudes $A_n^{(color)}$
of  color-dressed scalar theory considered in \cite{Du:2011js}. In that paper
 we have shown that amplitudes $A_n^{(color)}$ satisfy color-order
reversed relations, $U(1)$ decoupling relations, KK-relations
and both on-shell and off-shell BCJ relations. Because of the similarity
between amplitudes  $A_n^{(color)}$ and $A_{n;k}^*$,
it is natural to ask if the
$A_{n;k}^*$ defined by (\ref{A-k}) obey these same identities. We
can not make the naive conclusion  since there are differences between
these two theories. First the kinematic coupling constant $f^{ab}_{~~c}$ here
is only antisymmetric between $a,b$ while group structure constant
$f^{abc}$ of $U(N)$   is totally antisymmetric. In
addition, $f^{ab}_{~~c}$ depends on kinematics while $f^{abc}$ is independent
of momenta. Bearing these in mind, we  discuss properties of new amplitudes in this section.

\subsection{The color-order reversed relation}

\begin{figure}[!h]
  \centering
  \subfigure[reversing the color ordering in a five point graph]{
 \includegraphics[width=8cm]{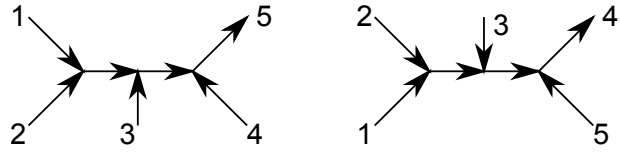}
 ~~~\label{fig-U1-a} }
 \\
 \subfigure[two typical terms in $U(1)$-decoupling relation]{
 \includegraphics[width=9cm]{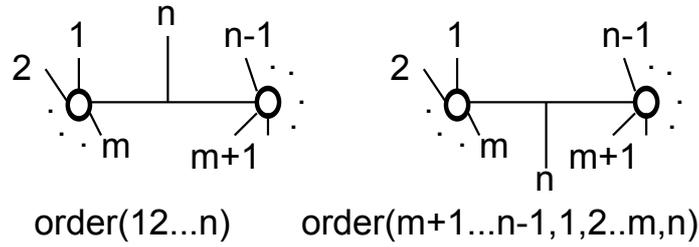}
 ~~~\label{fig-U1-b} }
 \caption{Demonstration of color-order reversed relation (part (a)) and the $U(1)$-decoupling relation
 (part (b)) }
\end{figure}

Since each $n_{i,k}^*$ is given by single graph, it is easy to
analyze it directly. Under the color-order reversing, each cubic vertex
will gain a minus sign coming from $f^{ab}_{~~c}=-f^{ba}_{~~c}$ (See
 Figure \ref{fig-U1-a} for example). For $n$-points amplitudes,
there are $(n-2)$ cubic vertices and $(n-3)$ propagators, thus we
will get a sign $(-)^{n-2}$, i.e., we do have
\bea A_{n;k}^* (123...n)=(-)^n
A_{n;k}^*(n...321)~.~~~~\Label{Ak-reverse}\eea
%

%

To see $U(1)$-decoupling relation
\bea \sum_{cyclic} A_{n;k}^*(C(1,2,...,n-1),n)=0
~~~\Label{A*-U1}\eea
is satisfied, we draw two typical terms in the cyclic sum in Figure
\ref{fig-U1-b}). These two terms have same denominator and same
numerator up to a sign since the  only difference between them is
the reversing of vertex connecting $n$, thus contributing $(-)$
sign. However, the left term belongs to color ordering $(123,...,n)$
while the right term belongs to color ordering
$(m+1,...,n-1,1,2,..,m,n)$, thus we can see the general
pair-by-pair cancellation in $U(1)$ identity given in (\ref{A*-U1}).

\subsection{The off-shell and on-shell BCJ relation}

Just like the color-dressed scalar field theory, the $A_{n;k}^*$  
satisfies a similar off-shell BCJ relation, which can be represented graphically by
\begin{equation}
\includegraphics[width=14cm]{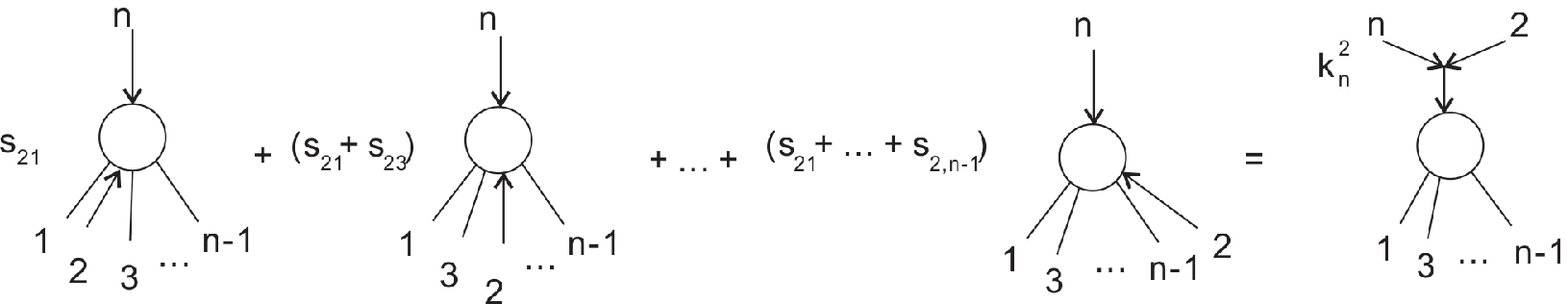}~~~\Label{eqn:1}
\end{equation}
with the momentum of particle $n$  taken off-shell. Depending on the
arrow directions of $2,n$ we have another two similar relations
\begin{equation}
\includegraphics[width=14cm]{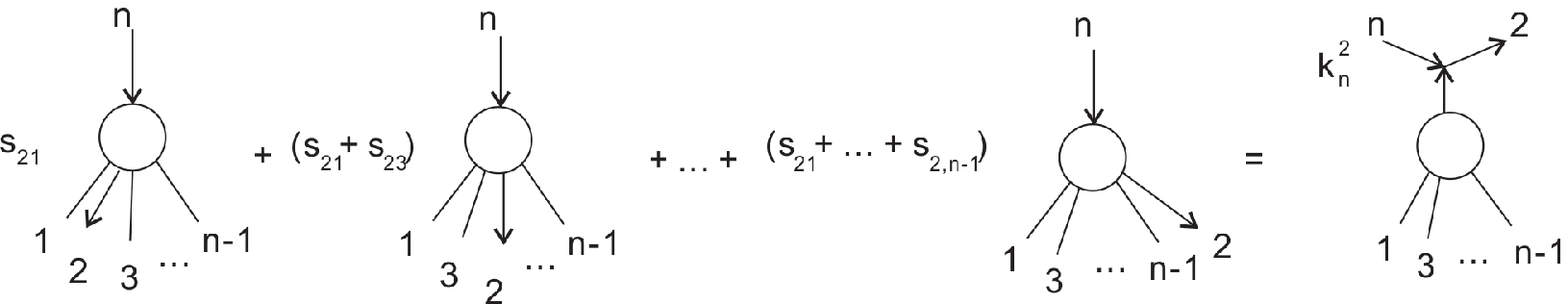}~~~\label{eqn:2}
\end{equation}
and
\begin{equation}
\includegraphics[width=14cm]{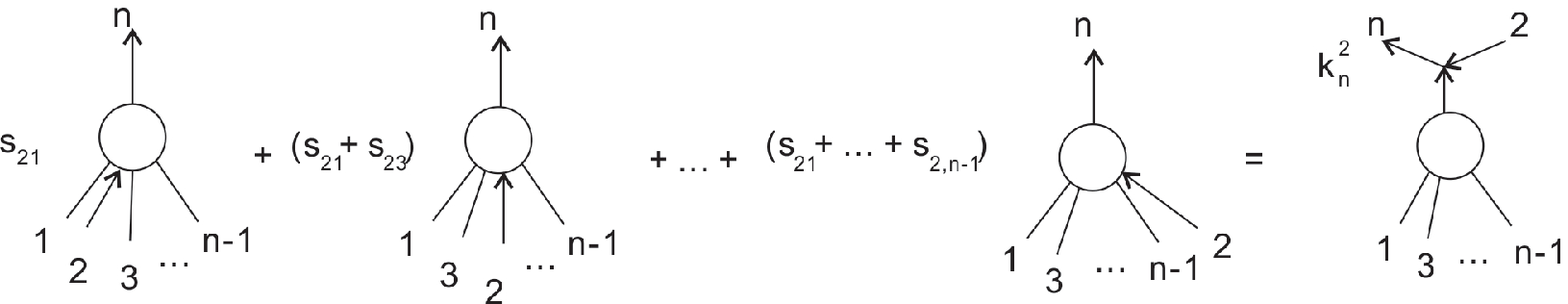}~~~\Label{eqn:3}
\end{equation}
There three relations  show that the off-shell BCJ relations are, as in
the case of color-order reversed and $U(1)$-decoupling relations,
independent of the choice of arrow directions. The proof of
relations (\ref{eqn:1}) is similar to the proof given in
\cite{Du:2011js} for color-dressed scalar theory.

The case of $n=3$ is trivially true from momentum conservation. For
$n=4$,  the left handed side of relation (\ref{eqn:1}) consists of 
following sum of graphs,
\bean
\includegraphics[width=15cm]{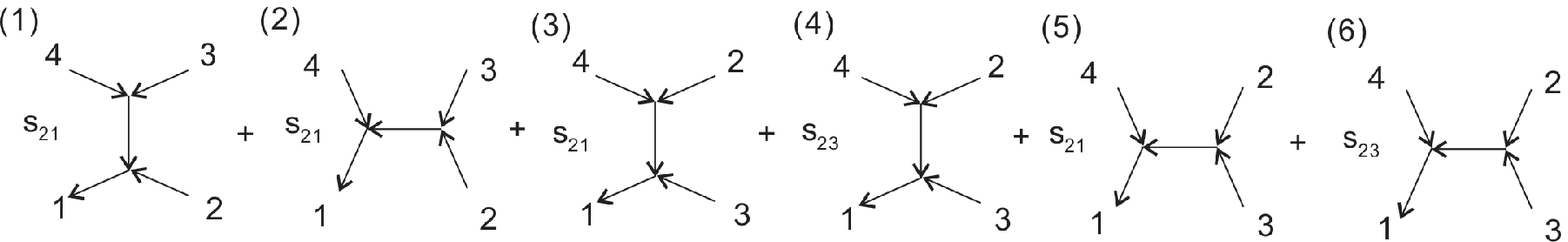}
\eean
We note that graphs (2) and (5) cancel due to antisymmetry of the
structure constant. Using Jacobi identity, graphs (1) and (6)
combine to produce
\bean
\includegraphics[width=8cm]{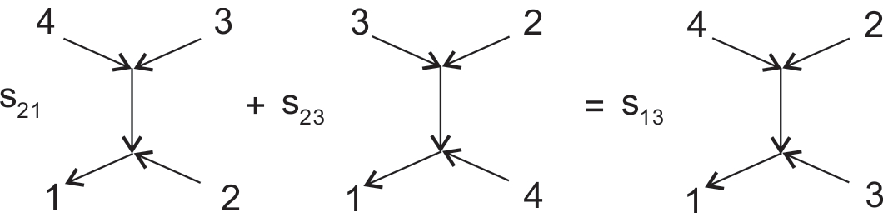} \eean
which, when  added to the rest two graphs (3) and (4), produces
the result as claimed using the on-shell conditions of
particles $1,2,3$.
\bean \includegraphics[width=7cm]{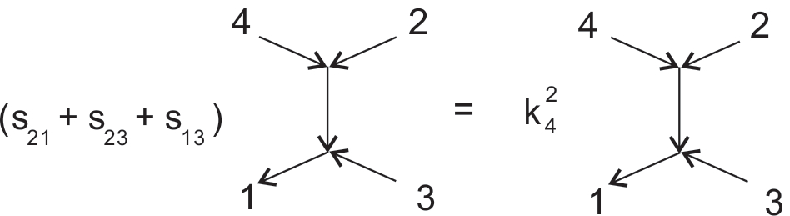} \eean
 Similar
manipulations can be done for (\ref{eqn:2}) and(\ref{eqn:3}).

Having proven  the $n=4$  example, let us  consider, for example,
relation (\ref{eqn:3}) for general $n$. We divide
contributions to any amplitude $A_{n;k}^*$ into the two sub-amplitudes
that share  same cubic vertex
with leg $n$. (See  part  Figure \ref{fig-U1-b} as an illustration.) i.e.,
\bea A_{n;k}^*=\sum_{ \#(n_L)=1}^{ \#(n_L)=n-2} A_{L}(\{ n_L\}; P_L)
V_3(n,P_L, P_R) A_R( -P_R; \{n_R\})~~~\Label{A*-split}\eea
where the number of legs in set $n_L$ can be $1,2,..,(n-2)$, and
 we used $V_3(n,P_L, P_R)$ to denote the cubic
vertex that connects leg $n$ to the two sub-amplitudes.
Using this decomposition,
the left handed side of
(\ref{eqn:3}) can be expressed by following graphs
\bea
\includegraphics[width=15cm]{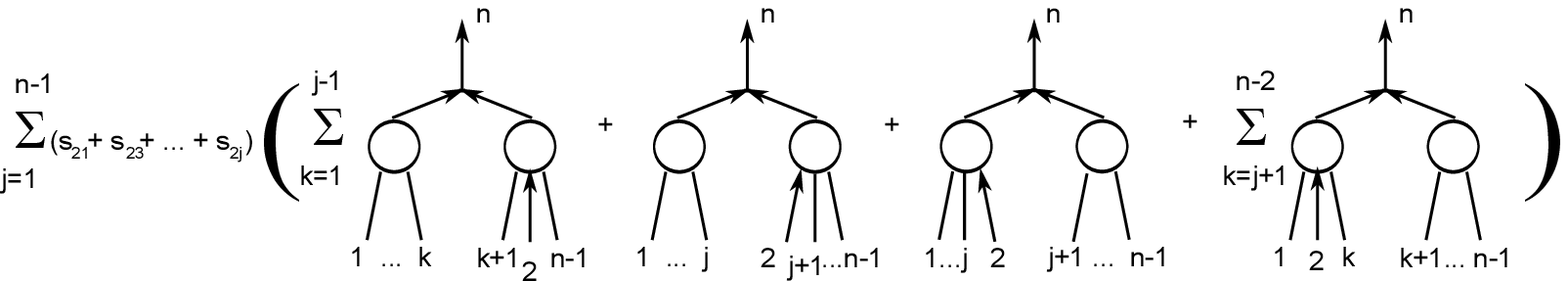}~~~\Label{npt1v}
\eea
We can categorize terms in (\ref{npt1v}) according to whether leg $2$ belongs
the left or right sub-amplitude. When the $2$ belongs to the left, the summation is
given by
\bea
\includegraphics[width=14cm]{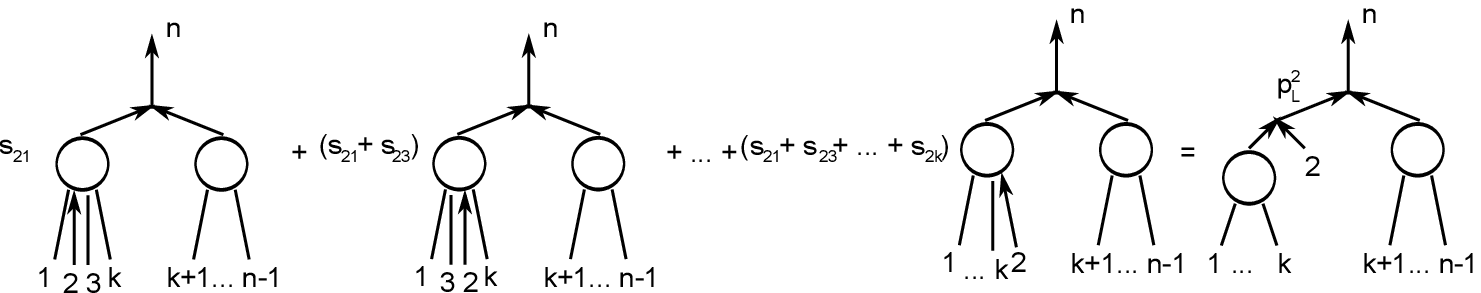}~~~\Label{npt2v}
\eea
where we used the  off-shell BCJ relation for left part with fewer points. The
value of $k$ in sum (\ref{npt2v}) can be $1,3,4,...,n-2$.
Similarly, when leg $2$ belongs to the right sub-amplitude, the summation is given by
\bea
\includegraphics[width=16cm]{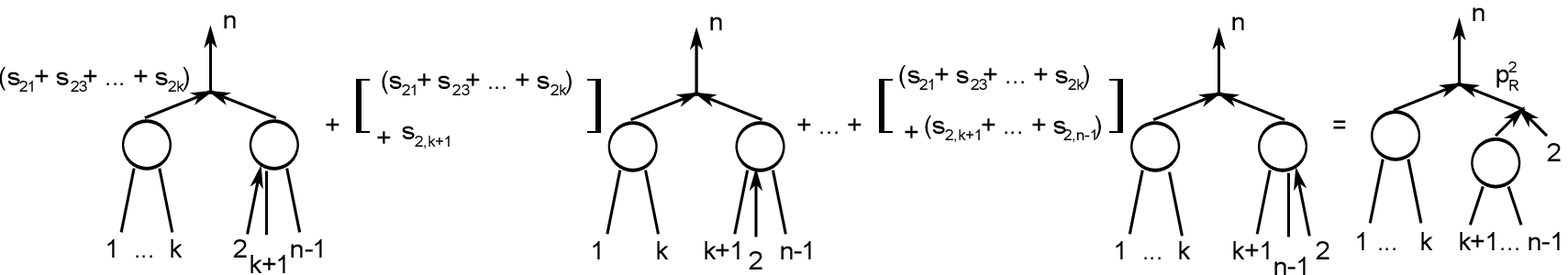}~~~\Label{npt3v}
\eea
The above sum can be split into  two parts. First there are terms
carrying the common factor $\sum_{i=1}^k s_{2i}$, and their sum
$(\sum_{i=1}^k s_{2i}) (A_R(2,k+1,...,n-1,P_R)+
A(k+1,2,...,n-1,P_R)+...+A(k+1,...,n-1,2,P_R))=0$ by
$U(1)$-decoupling identity. The remaining
part can be simplified by off-shell BCJ relation for fewer points.
The value of $k$ for sum (\ref{npt3v}) can be
$1,3,4,...,n-2$. The sum given in (\ref{npt2v}) and (\ref{npt3v})
can be further combined to
\bea
\includegraphics[width=8cm]{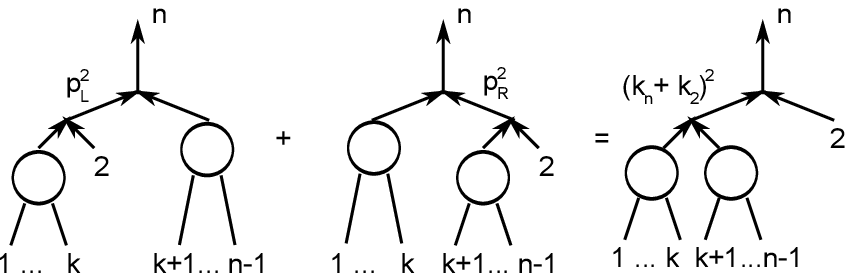}~~~\Label{npt4v}
\eea
by using the Jacobi identity derives from permuting the
internal $4$-point tree $\{n,2,
\{1,...,k\},\{k+1,...,n-1\}\}$. When we sum over $k$, we get
$(k_n+k_2)^2 V_3(P_L, 2, n) A_{n-1}^* (1,3,...,n-1, P_L)$. Finally,
result given in (\ref{npt4v}) is combined with term
$(\sum_{j=1}^{n-1} s_{2j}) V_3(P_L, 2, n) A_{n-1}^* (1,3,...,n-1,
P_L)$ coming from the decomposition of $A_n^*(1,3,4,...,n-1,2,n)$
according to (\ref{A*-split}) ( which is the boundary term that has
been neglected in the sum (\ref{npt2v}) and (\ref{npt3v})). Putting
together, we have the same graph multiplied by
$(k_n+k_2)^2- 2k_2\cdot k_n=k_n^2$, which is exactly
the right handed side of (\ref{eqn:3}). In other words, we have
proved the off-shell BCJ relation for $A_{n;k}^*$ amplitudes defined
in (\ref{A-k}). Taking the on-shell limit $k_n^2\to 0$,  we get the
familiar on-shell BCJ relation.

\subsection{The KK-relation}

The KK-relation found originally in \cite{Kleiss:1988ne} for gauge
theory is given by
\bea
A_n(\beta_1,...,\beta_r,1,\alpha_1,...,\alpha_s,n)=(-1)^{r}\Sl_{\{\sigma\}\in
P(O\{\alpha\}\cup O\{\beta\}^T)}A_n(1,\{\sigma\},n),~~\Label{KK}
 \eea
where the sum is over all permutations keeping relative ordering
inside the set $\a$ and the set $\b^T$ (where the $T$ means the
 set $\b$ with its order reversed), but at the same time
  allowing all relative
orderings between  sets $\a$ and $\b$. We  show that relation
(\ref{KK}) still holds if we replace $A_n$ by the fixed $k$ amplitude
$A_{n;k}^*$ defined in
(\ref{A-k}).

 When $\{\a\}$ is empty set (\ref{KK}) reduces to the color-order reversed
relation \eqref{Ak-reverse}, while  when there is only one leg
in the set $\{\beta\}$ or the set $\{\a\}$, (\ref{KK}) reduces to
the $U(1)$-decoupling identity \eqref{A*-U1}. Thus the color-order
reversed relation and the $U(1)$-decoupling identity are just two
special cases of KK relation. Since
KK-relations coincide with these two relations
for $n\leq 5$, the starting point of our
induction proof is checked.

Now we give the proof. Using the graphical representation, when the
set $\{\a\}$ is not empty, there are two types of graphs
depending on if $1$ is at the left or
right handed side of $n$. When leg $1$ is at the right
sub-amplitude
as described by the left graph of (\ref{KK-proof-1})
\begin{equation}
\includegraphics[width=12cm]{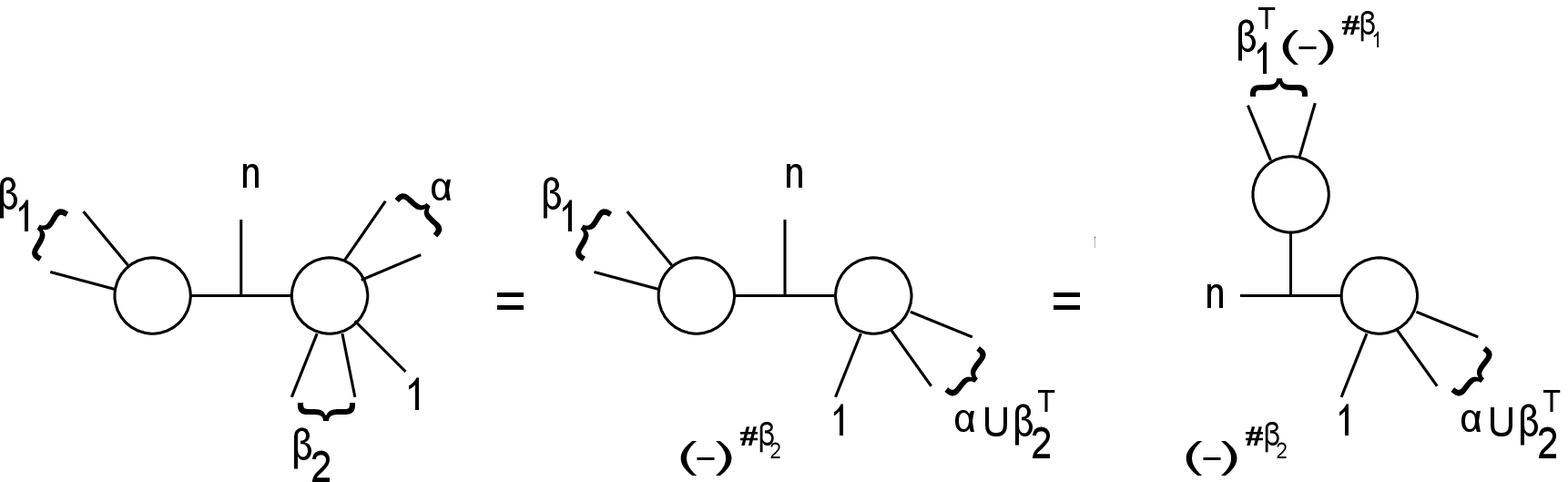}~~~\Label{KK-proof-1}
\end{equation}
we can  use  KK-relation of the right sub-amplitude to get the
middle graph of (\ref{KK-proof-1}). After that we  reverse the ordering of
sub-amplitude $\b_1$ and flip it to the  right hand side of $n$. The final result
is the
last graph of  (\ref{KK-proof-1}). When leg $1$ is at
the left handed side of $n$ as given  by
the left graph of  (\ref{KK-proof-2}),
\begin{equation}
\includegraphics[width=8cm]{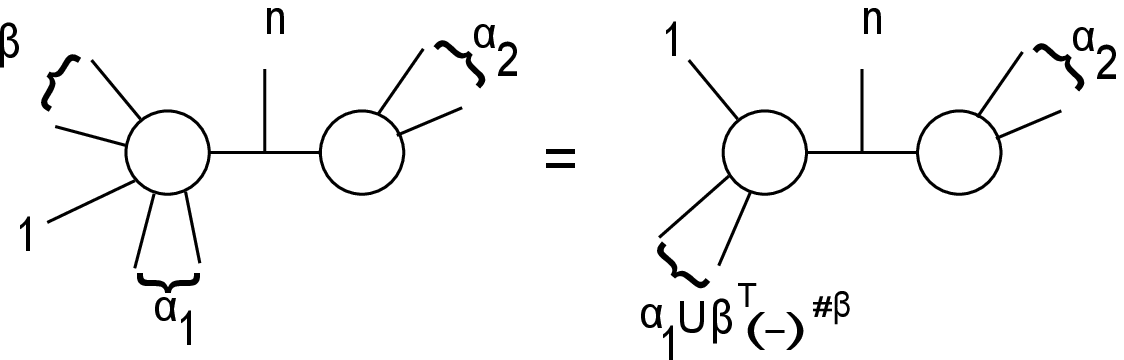}~~~\Label{KK-proof-2}
\end{equation}
we can use  KK-relation for the left sub-amplitude to get the right graph of (\ref{KK-proof-2}).
 When we combine 
results from (\ref{KK-proof-2}) and (\ref{KK-proof-1}),
we find they are nothing but  the graphical representation of right
handed side of equation (\ref{KK}) except contributions from following graphes
\begin{equation}
\includegraphics[height=3cm]{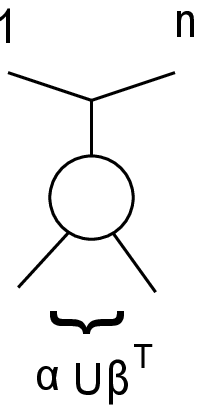}
\end{equation}
These contributions are nothing, but\footnote{Generalized $U(1)$-decoupling equation
(\ref{Two-KK}) has been written down in \cite{Berends:1988zn}.}
\bea V_3(n1P)\left(\sum\limits_{\{\sigma\}\in P(O\{\alpha\}\cup
O\{\beta^T\})}A^*_{n-1}(P_{1,n},\{\sigma\})\right)=0~.
\label{Two-KK}\eea
Sum inside the bracket of (\ref{Two-KK}) to be  zero can be 
proved by exactly same method as that given in
\cite{Du:2011js} after using two times of KK-relation for $(n-1)$-point
amplitudes.

\section{ Kinematic ordering of gauge theory amplitude}
\label{sec:DDM}

As proposed in \cite{Bern:1999bx} and proved in \cite{Du:2011js},
the full gauge theory amplitude can be represented by the
manifestly $(n-2)!$ symmetric KLT formula (which was found in
\cite{BjerrumBohr:2010ta})
\bea {\cal A}_n=(-)^n\sum_{\gamma,\b}{\W A(n,\gamma(2,...,n-1),1)
{\cal S}[ \gamma(2,...,n-1)|\b(2,..,n-1)]_{p_1}
A(1,\b(2,...,n-1),n)\over s_{123..(n-1)}}~~~\Label{newKLT}\eea
where leg $k_n$ has to be taken off-shell prior to the summation and
the full amplitude is given by the limit $k_n^2\to 0$.
The momentum kernel ${\cal
S}$ is defined as
\bea {\cal S}[i_1,...,i_k|j_1,j_2,...,j_k]_{p_1} & = & \prod_{t=1}^k
(s_{i_t 1}+\sum_{q>t}^k \theta(i_t,i_q) s_{i_t
i_q})~,~~~\Label{S-def}\eea
where $\theta(i_t, i_q)$ is zero when  pair $(i_t,i_q)$ has the same
ordering in both set ${\cal I},{\cal J}$ and otherwise it is one.
In the KLT formulation above, one copy of the  amplitudes $\W A$ is calculated from the
color-dressed scalar theory discussed in \cite{Du:2011js} and other
copy $A$ is  the familiar color-ordered gauge theory amplitude.

To calculate the sum in numerator of (\ref{newKLT}), let us consider
the following sum for given fixed ordering of $\gamma$, for example,
$\gamma(2,...,n-1)=(2,3,...,n-1)$,
\begin{equation}
\sum_{\{i\}\in S_{n-2}}\mathcal{S}[2,3,\dots n-1|i_{2},i_{3},\dots
i_{n-1}]_{k_1} \,{A}_{n}(1,\, i_{2},i_{3},\dots
i_{n-1};n)\Label{eq:offshellbcjx}\end{equation}
 where
the semicolon is used to emphasize that leg $n$ is  taken
off-shell. For amplitudes given by 
\bea A_n =\sum_{i=1}^N c_i \eps(q_i) \cdot ( \sum_{k=1}^n
A_{n;k}^*)~~~\Label{An-form}\eea
since  the part $\sum_{i=1}^N c_i \eps(q_i) $ is {\sl same for
all color orderings}, using the definition of function $\mathcal{S}$
we see that the sum in (\ref{eq:offshellbcjx}) can be written
as\footnote{In this form, we have used $V_3$, which is not exactly
right since we have not included the factor $\sum_{i=1}^N c_i
\eps(q_i)  $.}
\begin{eqnarray}
& \sum_{\{j\}\in S_{n-3}}\mathcal{S}[3,\dots n-1|j_{3},\dots
j_{n-1}]
\hspace{5cm} \nonumber \\
& \hspace{0.5cm}\times\left[s_{21}A_{n}(1,2,j_{3}\dots
j_{n-1};n)+(s_{21}+s_{2j_{3}})\tilde{A}_{n}(1,j_{3},2,\dots
j_{n-1};n)+\dots\right]
\nonumber \\
& ={p_{n}^{2}\over p_{n2}^2} V_{3}(2nc)\sum_{\{j\}\in
S_{n-3}}\mathcal{S}[3,\dots n-1|j_{3},\dots
j_{n-1}]\tilde{A}_{n-1}(1,j_{3},\dots;c).
\end{eqnarray}
where $\{ j \}$ is the set defined by deleting leg $2$ from the set
$\{ i \}$. In the last line we have used off-shell BCJ relation
(\ref{eqn:3}) as well as the form (\ref{An-form}). The sum over the new
${\cal S}$ can be done similarly and we obtain
\bea {p_{n}^{2}\over p_{n2}^2} V_3(2nc){p_{2n}^{2}\over p_{n23}^2}
V_3(3cc_1)\sum_{\{j\}\in S_{n-4}}\mathcal{S}[4,\dots n-1|j_{3},\dots
j_{n-1}]\tilde{A}_{n-1}(1,j_{4},\dots;c_1) \eea
Repeatedly reducing  the number of legs contained in the amplitude
one by one for $A_{n;k}^*$ we arrive at  the
graphical representation
\bea\includegraphics[width=13cm]{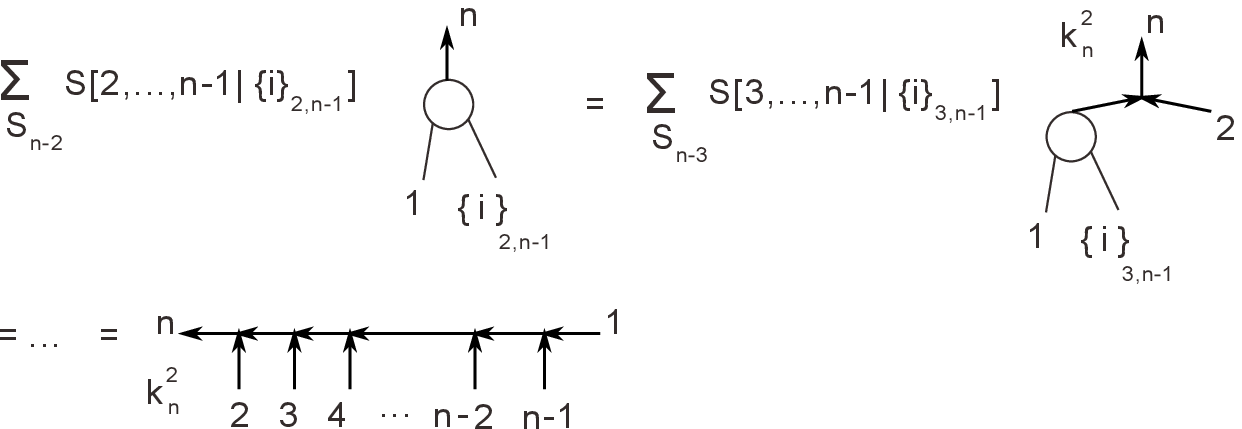} ~~~\Label{Ank-S-sum}\eea
Putting this result back to amplitudes given by (\ref{An-form}), the KLT formula
(\ref{newKLT}) produces naturally the following expression
\bea {\cal A}_n =\sum_{\gamma(23...(n-1))\in S_{n-2}}
\tilde{A}(1\gamma n)\sum_{j=1}^N c_j\eps(q_j)\cdot \left(
\begin{array}{c}
 \includegraphics[width=4cm]{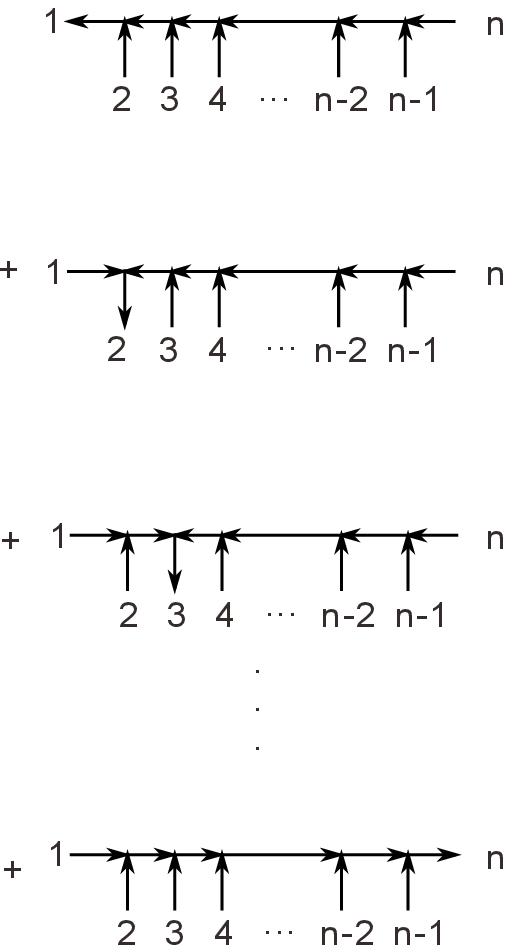} \\ \vspace{-1.5cm} \end{array} \right)~~~\Label{An-DDM}\eea
~\\
~\\
The graph at the right handed side of
(\ref{An-DDM}) is very similar to the chain of $U(N)$ group structure
constant  given in \cite{DelDuca:1999rs}. The manipulation
demonstrated above can obviously be applied  to KLT relation of gravity theory,
so  graviton amplitudes can  be ordered by
the same kinematic structure constants.

\section{Various forms of amplitudes}
\label{sec:formulations}

From recent progresses we saw that amplitudes of gauge theory
can be expressed in  following three formulations \cite{Bern:2008qj,DelDuca:1999rs}:
\bea {\rm double-copy~form}: &~~~ & {\cal A}_{tot} = \sum_i { c_i n_i\over
D_i}~~~\Label{BCJ-form}\\
{\rm Trace~form}: &~~~ &{\cal A}_{tot}  = \sum_{\sigma\in
S_{n-1}}{\rm Tr} (T^{\sigma_1}... T^{\sigma_n})
A(\sigma)~~~\Label{Trace-form}\\
{\rm DDM~form}: &~~~ & {\cal A}_{tot}  =  \sum_{ \sigma\in S_{n-2}}
c_{1|\sigma(2,..,n-1)|n} A(1,\sigma,n)~~~\Label{DDM-form} \eea
where $A$ are color ordered amplitudes, $T^a$ is the matrix of
fundamental representation of $U(N)$ group and
$c_i,c_{1|\sigma(2,..,n-1)|n} $ are constructed using the structure
constants $f^{abc}$. For example, we have
\bea c_{1|\sigma(2,..,n-1)|n}=f^{\sigma_1 \sigma_2 x_1} f^{x_1
\sigma_3 x_2}... f^{x_{n-3} \sigma_{n-1} n}~~~\Label{DDM-c}\eea
The transformation from double-copy formulation to  DDM was shown in
\cite{Du:2011js} using the KLT relation, while the transformation
from DDM to Trace was given in \cite{DelDuca:1999rs} where
the following two properties of Lie algebra of $U(N)$ gauge group
were essential
\bea {\rm Property~One:} & ~~~ & (f^a)_{ij}= f^{aij}={\rm Tr}( T^a[
T^i,
T^j]),~~~\Label{group-1}\\
{\rm Property~Two:} & ~~~ &  \sum_a {\rm Tr}( X T^a) {\rm Tr}(T^a
Y)= {\rm Tr}( XY)~~\Label{group-2}\eea
A special feature of double-copy formulation is that
both $c_i$ and $n_i$ satisfy
the Jacobi identity in corresponding Feynman diagrams with only
cubic vertexes. Because of this duality, it is natural to exchange
the role between $c_i$ and $n_i$ and consider the following two dual
formulations
\bea {\rm Dual~Trace~form}: &~~~ &{\cal A}_{tot}  = \sum_{\sigma\in
S_{n-1}} \tau_{\sigma_1... \sigma_n}
\W A(\sigma)~~~\Label{Dual-Trace-form}\\
{\rm Dual~DDM~form}: &~~~ & {\cal A}_{tot}  =  \sum_{ \sigma\in
S_{n-2}} n_{1|\sigma(2,..,n-1)|n}\W
A(1,\sigma,n)~~~\Label{dual-DDM-form} \eea
where $\W A$ is color ordered scalar theory with $f^{abc}$ as cubic
coupling constants (see the references    \cite{Bern:1999bx,
Du:2011js}) and $\tau$ is required to be cyclic invariant. Indeed,
the Dual-DDM was given in \cite{Bern:2010yg}  while the
Dual-Trace-form was conjectured in \cite{Bern:2011ia} with
explicit constructions given for the first few lower-point amplitudes
and a general
construction was given in \cite{BjerrumBohr:2012mg}.  Although the
existence of above two dual formulations were established,
a systematic
Feynman rule-like prescription to the coefficients $\tau$ and $n$ is
not known at this moment. Our result (\ref{An-DDM})  $n_{1\sigma n}$ for dual-DDM-form 
serves as a
small step towards this goal.

Having above explanation, let us consider following situation
where both $c_i, n_i$ satisfying Jacobi-identity can be constructed by Feynman rule,
i.e.,  the theory can be constructed using cubic vertex with
coupling constant ${\cal F}_{abc} \W {\cal F}_{abc}$. We want to know
under this assumption, which dual form comes out naturally. The conclusion we found is that
the dual DDM (\ref{dual-DDM-form}) is more
compatible with double-copy formulation.

\begin{figure}
  \centering
 \includegraphics[width=1\textwidth]{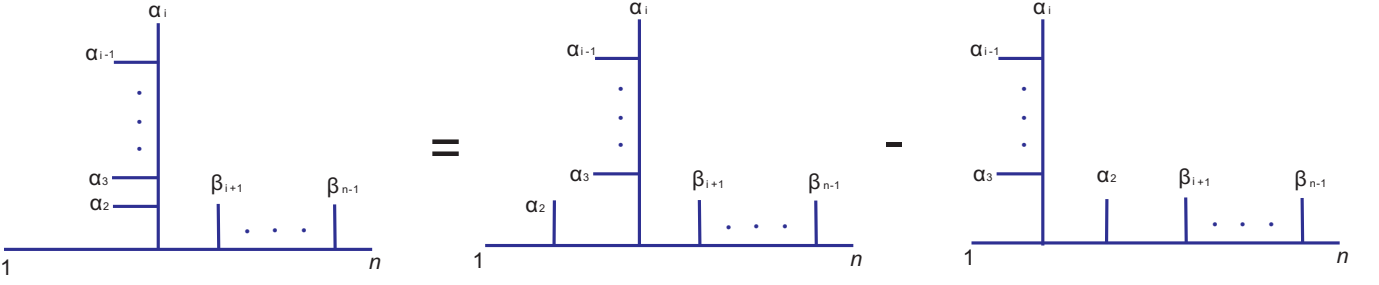}
 \caption{We can use Jacobi identity to reduce the contraction of $\mathcal{F}$s.}
 \label{Jacobi-decomposition}
\end{figure}

\begin{figure}
  \centering
 \includegraphics[width=0.4\textwidth]{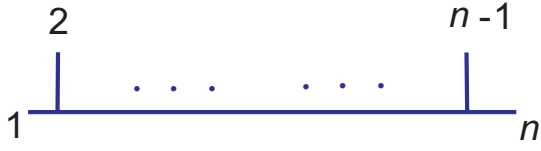}
 \caption{A DDM chain with contractions of structure constants $\mathcal{F}^{12e_1}
 \mathcal{F}^{e_13e_2}...\mathcal{F}^{e_{n-3},n-1,n}$} \label{DDF}
\end{figure}

To see that, let us note that the total
amplitude can be constructed recursively as
\bea {\cal A}(1,2,...,n)
&=&\Sl_{i=1}^{n-1}\Sl_{Split}\mathcal{F}^{1e_1e_2}
\widetilde{\mathcal{F}}^{\bar 1\bar e_1\bar e_2}\frac{{\cal
A}(e_1,u_1,...,u_i)}{P_{u_1,...,u_i}^2}\frac{{\cal
A}(e_2,v_{1},...,v_{n-2-i},n)}
{P_{v_1,...,v_{n-2-i},n}^2},~~\Label{rec-I} \eea
where the second sum is over all possible separations of $(n-1)$
particles into two subsets $\{ u\}, \{v\}$ with $n_u=i$. Assuming
the color-decomposition holds for lower-point amplitude ${\cal A}$,
we can substitute the lower-point DDM-form  into above equation and
obtain
\bea {\cal A}(1,2,...,n)
&=&\Sl_{i=1}^{n-1}\Sl_{Split}\mathcal{F}^{1e_1e_i}\widetilde{\mathcal{F}}^{\bar
1\bar e_1\bar e_i}\times \left(\Sl_{\alpha\in
perm\{u_1,...,u_{i-1}\}}\mathcal{F}^{e_1\alpha_1
e_2}\mathcal{F}^{e_2\alpha_2 e_3}...\mathcal{F}^{e_{i-1}\alpha_{i-1}
u_i} \frac{\W
A(e_1,\alpha_1,...,\alpha_{i-1},u_i)}{P_{u_1,...,u_i}^2}\right)\nn
&&\times\left(\Sl_{\beta\in
perm\{v_1,...,v_{n-2-i}\}}\mathcal{F}^{e_i\beta_1
e_{i+1}}\mathcal{F}^{e_{i+1}\beta_2 e_{i+2}}...\mathcal{F}^{
e_{n-3}\beta_{n-i-2} n} \frac{\W
A(e_i,\beta_1,...,\beta_{n-i-2},n)}{P_{v_1,...,v_{n-i-2},n}^2}\right).
\nn
& = & \Sl_{i=1}^{n-1}\Sl_{Split}
\mathcal{F}^{1e_1e_i}\mathcal{F}^{e_1\alpha_1
e_2}...\mathcal{F}^{e_{i-1}\alpha_{i-1} u_i} \mathcal{F}^{e_i\beta_1
e_{i+1}}...\mathcal{F}^{ e_{n-3}\beta_{n-i-2} n}\nn & &
\left[\widetilde{\mathcal{F}}^{\bar 1\bar e_1\bar e_i}\frac{\W
A(e_1,\alpha_1,...,\alpha_{i-1},u_i)}{P_{u_1,...,u_i}^2} \frac{\W
A(e_i,\beta_1,...,\beta_{n-i-2},n)}{P_{v_1,...,v_{n-i-2},n}^2}\right]~~\Label{rec-1}
\eea
where for given permutations $\alpha_1$,...,$\alpha_i$ of $u$s and
$\beta_1$,...,$\beta_{n-i-2}$ of $v$s, the contraction of
$\mathcal{F}$s has the structure at the left handed side of  Fig.
\ref{Jacobi-decomposition}. After  applying Jacobi identity,
$\mathcal{F}^{1e_1e_i}\mathcal{F}^{e_1\alpha_1  e_2}$ becomes
\bea \mathcal{F}^{1e_1e_i}\mathcal{F}^{e_1\alpha_1
e_2}=\mathcal{F}^{1\alpha_1 e_1}\mathcal{F}^{e_1e_2e_i}
-\mathcal{F}^{1e_2e_1}\mathcal{F}^{e_1\alpha_1 e_i}, \eea
i.e., the right handed side of  Fig. \ref{Jacobi-decomposition}.
Iterating this procedure like the one did in \cite{DelDuca:1999rs},
we get a sum of $2^{i-1}$ DDM chains (e.g., Fig. \ref{DDF}) where
the ordered set $O\{\alpha_1,...,\alpha_{i-1}\}$ is split into two
ordered sets $O\{\sigma\}$ and $O\{\rho\}$ and  the form is  given
by
$(-1)^s\mathcal{F}^{1\sigma_1e}...\mathcal{F}^{e\sigma_{t}e}\mathcal{F}^{e
u_i
e}\mathcal{F}^{e\rho_se}...\mathcal{F}^{e\rho_1e}\mathcal{F}^{e\beta_1e}
...\mathcal{F}^{e_{n-3},\beta_{n-2-i},n}$. All these forms are
multiplied by $\widetilde{\mathcal{F}}^{\bar 1\bar e_1\bar e_i}\W
A(e_1,\alpha_1,...,\alpha_{i-1},u_i)\W
A(e_i,\beta_1,...,\beta_{n-i-2},n)$. Doing same things to other
permutations of $u_1,...,u_{i-1}$s and collecting all terms having
same DDM chain structure, we get
\bea
&&\mathcal{F}^{1\sigma_1e_1}...\mathcal{F}^{e\sigma_{t}e}\mathcal{F}^{e
u_i
e}\mathcal{F}^{e\rho_se}...\mathcal{F}^{e\rho_1e}\mathcal{F}^{e\beta_1e}...
\mathcal{F}^{e_{n-3},\beta_{n-2-i},n}\nn
&\times&\widetilde{\mathcal{F}}^{\bar 1\bar e_1\bar
e_i}\frac{1}{P_{u_1,...,u_i}^2}\left[\Sl_{\gamma\in
OP(\{\sigma\}\bigcup\{\rho\})}\W
A(e_1,\gamma_1,...,\gamma_{i-1},u_i)\right]\frac{1}{P_{v_1,...,v_{n-2-i},n}^2}\W
A(e_i,\beta_1,...,\beta_{n-i-2},n)\nn
&=&\mathcal{F}^{1\sigma_1e_1}...\mathcal{F}^{e\sigma_{t}e}\mathcal{F}^{e
u_i
e}\mathcal{F}^{e\rho_se}...\mathcal{F}^{e\rho_1e}\mathcal{F}^{e\beta_1e}...
\mathcal{F}^{e_{n-3},\beta_{n-2-i},n}\nn
&\times&\widetilde{\mathcal{F}}^{\bar 1\bar e_1\bar
e_i}\frac{1}{P_{u_1,...,u_i}^2}\W
A(e_1,\sigma_1,...,\sigma_t,u_i,\rho_s,...,\rho_1)
\frac{1}{P_{v_1,...,v_{n-2-i},n}^2}\W
A(e_i,\beta_1,...,\beta_{n-i-2},n). \eea
where the KK-relation has been used for the sum in square bracket.

Putting this result back to recursion relation we reach our final
claim
\bea {\cal A}(1,2,...,n) &=&\Sl_{i=1}^{n-1}\Sl_{Split}
\Sl_{\alpha\in S\{u\}}\Sl_{\beta\in
S\{v_1\}}\Bigl[\mathcal{F}^{1\sigma_1e_1}...\mathcal{F}^{e\sigma_{t}e}
\mathcal{F}^{e u_i
e}\mathcal{F}^{e\rho_se}...\mathcal{F}^{e\rho_1e}\mathcal{F}^{e\beta_1e}...
\mathcal{F}^{e_{n-3},\beta_{n-2-i},n}\nn
&\times&\widetilde{\mathcal{F}}^{\bar 1\bar e_1\bar
e_i}\frac{1}{P_{u_1,...,u_i}^2}\W
A(e_1,\sigma_1,...,\sigma_t,u_i,\rho_s,...,\rho_1)
\frac{1}{P_{v_1,...,v_{n-2-i},n}^2}\W
A(e_i,\beta_1,...,\beta_{n-i-2},n)\Bigr]\nn
&=&\Sl_{\sigma\in
S_{n-2}}\mathcal{F}^{a_1a_{\sigma_2}e_1}...\mathcal{F}^{e_{n-3}
a_{\sigma_{n-1}}a_n}\times\Sl_{i=1}^{n-1}{\widetilde{\mathcal{F}}^{\bar
1\bar e_1\bar e_i}} \frac{\W
A(e_1,\sigma_1,...,\sigma_t,u_i,\rho_s,...,\rho_1)}{P_{u_1,...,u_i}^2}\frac{\W
A(e_i,\beta_1,...,\beta_{n-i-2},n)}{P_{v_1,...,v_{n-2-i},n}^2}\nn
&= & \Sl_{\sigma\in
S_{n-2}}\mathcal{F}^{a_1a_{\sigma_2}e_1}...\mathcal{F}^{e_{n-3}
a_{\sigma_{n-1}}a_n}\W A(1\sigma(2...n-1) n)
\eea
where at the last step we have used the recursion relation for color
ordered amplitudes.

\section{Conclusion}
In this paper we have presented an algorithm which allows systematic
construction of the BCJ numerators as well as the kinematic-dual to
the DDM formulation. We have shown that assuming gauge symmetry
provides enough degrees of freedom, we can express
tree-level amplitudes as linear combinations of cubic graph contributions,
where Jacobi-like relations between kinematic numerators
can be made manifest.

Although our construction is systematically, it is a little bit hard to
use practically. In other words, our results is just a small step toward
the simple construction of BCJ numerators, which can have important applications
for loop calculations of gravity amplitudes. 

\subsection*{Acknowledgements}

%
Y.J.Du would like to thank Profs. Yong-Shi Wu and Yi-Xin Chen for
helpful suggestions. He would also like to thank Qian Ma, Gang Chen,
Hui Luo, Congkao Wen and Yin Jia for helpful discussions. Y. J. Du
is supported in part by the NSF of China Grant No.11105118. CF is
grateful for Gang Chen, Konstantin Savvidy and Yihong Wang for
helpful discussions. Part of this work was done in Zhejiang
University and Nanjing University. CF would also like to acknowledge
the supported from National Science Council, 50 billions project of
Ministry of Education and National Center for Theoretical Science,
Taiwan, Republic of China as well as the support from S.T. Yau
center of National Chiao Tung University. B.F is supported, in part,
by fund from Qiu-Shi and Chinese NSF funding under contract
No.11031005, No.11135006, No. 11125523.

\appendix

\section{Off-shell KK relation from Berends-Giele recursion}

The KK-relation was first written down in \cite{Kleiss:1988ne}
without proof. With our knowledge, a proof can be found in
\cite{DelDuca:1999rs}. Since the off-shell tensors can be
constructed  by Berends-Giele recursion relation
\cite{Berends:1987me}, it is natural to prove the off-shell KK
relation by this recursion relation and in this appendix we provide
a proof for reader's convenience.

The off-shell KK relation is given as
\bea J(1,\{\alpha\},n,\{\beta\})=(-1)^{n_{\beta}}\Sl_{\sigma\in
OP(\{\alpha\}\bigcup\{\beta^T\})}J(1,\sigma,n),
~~~\Label{off-KK-BG}\eea
where  $J(1,2,...,n)$ is an off-shell tensor. After contracting $J$
with on-shell polarization vectors of external legs, it becomes a
color-ordered amplitude $A(1,2,...,n)$, and thus the off-shell KK
relation becomes the on-shell KK relation.

According to Berends-Giele recursion relation, for a given tensor,
we can pick out a leg, for example, the leg $1$, to construct whole
tensor recursively. In the formula, the leg $1$ can be connected to
either a three-point vertex or a four point vertex, i.e., we can
separate the tensor into
$J(1,2,...,n)=J^{(3)}(1,2,...,n)+J^{(4)}(1,2,...,n)$. We will do the
same separation at both sides of (\ref{off-KK-BG}) and show the
matching for each part.

 {\bf Connecting to 3-point vertex:}  In this case, the R.H.S. of KK
relation (\ref{off-KK-BG}) can be expressed by
\bea & & (-)^{n_\b}\Sl_{
                          \begin{array}{c}
                           \alpha\rightarrow\alpha_A,\alpha_B \\
                            \beta\rightarrow\beta_A,\beta_B \\
                          \end{array}
} \Sl_{
       \begin{array}{c}
         \sigma_A\in OP(\{\alpha_A\}\bigcup \{\beta_B\}^T) \\
        \sigma_B\in OP(\{\alpha_B\}\bigcup \{\beta_A\}^T) \\
       \end{array}
}
V_{(3)}^{1e_1e_2}\frac{1}{P_{\alpha_A,\beta_B}^2}J(e_1,\sigma_A)\frac{1}
{P_{\alpha_B,\beta_A}^2}J(e_2,\sigma_B,n),\nn
&=&(-)^{n_\b}\Sl_{
                          ~\alpha\rightarrow\alpha_A,\alpha_B;
                            \beta\rightarrow\beta_A,\beta_B
}
V^{1e_1e_2}_{(3)}\frac{1}{P_{\alpha_A,\beta_B}^2}\left(\Sl_{\sigma_A\in
OP(\{\alpha_A\}\bigcup\{\beta_B\}^T)}J(e_1,\sigma_A)\right)\nn
&&\times\frac{1}{P_{\alpha_B,\beta_A}^2}\left(\Sl_{\sigma_B\in
OP(\{\alpha_B\}\bigcup\{\beta_A\}^T)}J(e_2,\sigma_B,n)\right)~~~\Label{KK-3-sum}
\eea
where the first sum is over all possible splitting of set $\a,\b$
into two subsets (including the case, for example,
$\alpha_A=\emptyset$) and the second sum is over all possible
relative ordering between subsets $\a_i, \b_j$. Now we consider the
sum in (\ref{KK-3-sum}) for different splitting:
\begin{itemize}

\item {\bf{(i)}}\emph{ If both $\alpha_A$ and $\beta_B$ sets are
nonempty}, we can use lower-point generalized $U(1)$-decoupling
identity  (\ref{Two-KK})
 \bea
 \Sl_{\sigma_A\in
 OP(\{\alpha_A\}\bigcup\{\beta_B\}^T)}J(e_1,\sigma_A)=0.~~~\Label{A-3}
 \eea
 Thus this case does not have nonzero contribution.

\item  {\bf{(ii)}}  {\emph{If ${\beta_B}$ set is empty}}, we have \bea
\Sl_{\alpha\rightarrow\alpha_A,\alpha_B }
V^{1e_1e_2}\frac{1}{P_{\alpha_A}^2}J(e_1,\alpha_A)
\times\frac{1}{P_{\alpha_B,\beta}^2}J(e_2,\alpha_B,n,\beta), \eea
where we have used lower-point KK relation to sum up the last line
in (\ref{KK-3-sum}).

\item {\bf{(iii)}} \emph{If ${\alpha_A}$ is empty}, we have
\bea \Sl_{\beta\rightarrow\beta_A,\beta_B}
V^{1e_2e_1}\frac{1}{P_{\alpha,\beta_A}^2}J(e_2,\alpha,n,\beta_A)\times\frac{1}{P_{\beta_B}^2}J(e_1,\beta_B)
, \eea
where we have used lower-point KK relations for the second bracket,
the color-order reversed relation for the first brackets  as well as
the antisymmetry of three-point vertex $V^{1,2,3}=(-1)V^{1,3,2}$ (so
the overall factor $(-)^{n_\b}$ disappears).

The sum of  contributions from (ii) and (iii) is just the recursive
expansion of  $J^{(3)}(1,\alpha,n,\beta)$.

\end{itemize}

 {\bf Connecting to 4-point vertex:}    In
this case, the R.H.S. of KK relation (\ref{off-KK-BG}) is given as
\bea & & (-)^{n_\b}\Sl_{
         \alpha\rightarrow\alpha_A,\alpha_B,\alpha_C;
         \beta\rightarrow\beta_A,\beta_B,\beta_C}V_{(4)}^{1 e_1 e_2 e_3}
         \frac{1}{P^2_{\alpha_A,\beta_A}}\left(\Sl_{\sigma_A\in
OP(\{\alpha_A\}\bigcup\{\beta_A^T\})}J(e_1,\sigma_A)\right)\nn
&&\times\frac{1}{P^2_{\alpha_B,\beta_B}}\left(\Sl_{\sigma_B\in
OP(\{\alpha_B\}\bigcup\{\beta_B^T\})}J(e_2,\sigma_B)\right)\times\frac{1}{P^2_{\alpha_C,\beta_C}}
\left(\Sl_{\sigma_C\in
OP(\{\alpha_C\}\bigcup\{\beta_C^T\})}J(e_3,\sigma_C,n)\right).~~~\Label{A-6}
\eea
where the sum is over all possible splitting of sets $\a,\b$ into
three subsets (with possible empty subset). For given splittings
$\alpha\rightarrow \alpha_A,\alpha_B,\alpha_C$,
$\beta\rightarrow\beta_A,\beta_B,\beta_C$, there are several cases:

\begin{itemize}

\item {\bf(i)} \emph{If both $\{\alpha_A\}$ and $\{\beta_A\}$ are
nonempty or both $\{\alpha_B\}$ and $\{\beta_B\}$ are nonempty}, we
can use lower-point generalized $U(1)$-decoupling identity
(\ref{A-3}) and the sum is zero for the first or the second brackets
in (\ref{A-6}).

\item {\bf(ii)}\emph{If $\sigma_A=OP(\{\alpha_A\})$,
$\sigma_B=OP(\{\alpha_B\})$, $\sigma_C\in
OP(\{\alpha_C\}\bigcup\{\beta^T\})$ }, we have nonzero contribution
\bea \Sl_{\alpha\rightarrow\alpha_A,\alpha_B,\alpha_C
}V_{(4)}^{1e_1e_2e_3}&&\frac{1}{P^2_{\alpha_A}}J(e_1,\alpha_A)\frac{1}
{P^2_{\alpha_B}}J(e_2,\alpha_B)\frac{1}{P^2_{\alpha_C,\beta}}
J(e_3,\alpha_C,n,\beta), \eea
where we have used lower-point KK relation to sum up the last
bracket.

\item {\bf(iii)}\emph{If
$\sigma_A=OP(\{\beta^T_C\})$,$\sigma_B=OP(\{\beta^T_B\})$,
$\sigma_C\in OP(\{\alpha\}\bigcup\{\beta^T_A\})$}, we have nonzero
contribution \bea \Sl_{\beta\rightarrow\beta_A,\beta_B,\beta_C
}V_{(4)}^{1e_1e_2e_3}&&\frac{1}{P^2_{\alpha,\beta_A}}J(e_1,\alpha,n,\beta_A)\frac{1}
{P^2_{\beta_B}}J(e_2,\beta_B)\frac{1}{P^2_{\beta_C}}J(e_3,\beta_C),
\eea
where we have used lower-point KK relation for the third bracket and
 the color-order reversed relation for the first and second brackets as well as
the symmetry of four-vertex $V_{(4)}^{1234}=V_{(4)}^{1432}$.

\item {\bf(iv)}\emph{If $\sigma_A=OP(\{\alpha_A\})$,
$\sigma_B=OP(\{\beta^T_B\})$, $\sigma_C\in
OP(\{\alpha_B\}\bigcup\{\beta_A^T\})$}, the nonzero contribution is
given as
\bea (-)^{n_\b}\Sl_{
         \alpha\rightarrow\alpha_A,\alpha_B;
         \beta\rightarrow\beta_A,\beta_B
}V_{(4)}^{1e_1e_2e_3}&&\frac{1}{P^2_{\alpha_A}}J(e_1,\alpha_A)\frac{1}
{P^2_{\beta_B}}J(e_2,\beta^T_B)\frac{1}{P^2_{\alpha_B,\beta_A}}
\left(\Sl_{\sigma\in
OP(\{\alpha_B\}\bigcup\{\beta_A^T\})}J(e_3,\sigma,n)\right).\nn \eea
Similarly, \emph{If $\sigma_A=OP(\{\beta^T_B\})$,
$\sigma_B=OP(\{\alpha_A\})$, $\sigma_C\in
OP(\{\alpha_B\}\bigcup\{\beta_A^T\})$ }, we have
\bea (-)^{n_\b}\Sl_{
         \alpha\rightarrow\alpha_A,\alpha_B;
         \beta\rightarrow\beta_A,\beta_B
}V_{(4)}^{1e_2e_1e_3}&&\frac{1}{P^2_{\beta_B}}J(e_2,\beta^T_B)\frac{1}
{P^2_{\alpha_A}}J(e_1,\alpha_A)\frac{1}{P^2_{\alpha_B,\beta_A}}
\left(\Sl_{\sigma\in
OP(\{\alpha_B\}\bigcup\{\beta_A^T\})}J(e_3,\sigma,n)\right).\nn \eea
where it is worth to notice that the 4-point vertex is written as
$V_{(4)}^{1e_2e_1e_3}$. The reason doing so is because the 4-point
vertex is
\bea
V^{1234}=i\eta_{\mu_1\mu_3}\eta_{\mu_2\mu_4}-\frac{i}{2}(\eta_{\mu_1\mu_2}\eta_{\mu_3\mu_4}
+\eta_{\mu_1\mu_4}\eta_{\mu_2\mu_3}).
\eea
so we have following identity
\bea V^{1234}+V^{1324}=-V^{1243}. \eea
Using this identity and lower-point KK relation for the third
brackets and the color order reversed relation for the first or the
second brackets (thus the factor $(-)^{n_\b}$ disappears), the sum
of above two contributions becomes
\bea \Sl_{
         \alpha\rightarrow\alpha_A,\alpha_B;
         \beta\rightarrow\beta_A,\beta_B
}V_{(4)}^{1e_1e_3e_2}&&\frac{1}{P^2_{\alpha_A}}J(e_1,\alpha_A)\frac{1}{P^2_{\alpha_B,\beta_A}}
J(e_3,\alpha_B,n,\beta_A)\frac{1}{P^2_{\beta_B}}J(e_2,\beta_B).\nn
\eea
The sum of $(ii)$, $(iii)$, $(iv)$  is just
$J^{(4)}(1,\alpha,n,\beta)$.

\end{itemize}

Having shown both 3-point vertex part and 4-point vertex part have
KK-relation, we have shown the whole off-shell tensor
$J(1,\alpha,n,\beta)$ has the KK-relation. In the proof, we have
used  the antisymmetry of three-point vertex under exchanging a pair
of indices as well as the identity between 4-point vertex. This
proof shows that if a tensor is constructed only by  three-point
vertices, it obeys KK relation when the three-point vertex is
antisymmetry under exchanging a pair of indices.



\begin{thebibliography}{References}



\bibitem{Bern:2008qj}
  Z.~Bern, J.~J.~M.~Carrasco and H.~Johansson,
  ``New Relations for Gauge-Theory Amplitudes,''
  Phys.\ Rev.\ D {\bf 78}  (2008) 085011
  [arXiv:0805.3993 [hep-ph]].

\bibitem{BjerrumBohr:2009rd}
  N.~E.~J.~Bjerrum-Bohr, P.~H.~Damgaard and P.~Vanhove,
  ``Minimal Basis for Gauge Theory Amplitudes,''
  Phys.\ Rev.\ Lett.\  {\bf 103}, 161602 (2009)
  [arXiv:0907.1425 [hep-th]].


\bibitem{Stieberger:2009hq}
  S.~Stieberger,
  ``Open \& Closed vs. Pure Open String Disk Amplitudes,''
  arXiv:0907.2211 [hep-th].



\bibitem{BjerrumBohr:2010zs}
  N.~E.~J.~Bjerrum-Bohr, P.~H.~Damgaard, T.~Sondergaard and P.~Vanhove,
  ``Monodromy and Jacobi-like Relations for Color-Ordered Amplitudes,''
  JHEP {\bf 1006} (2010) 003
  [arXiv:1003.2403 [hep-th]].

\bibitem{Mafra:2011kj}
  C.~R.~Mafra, O.~Schlotterer and S.~Stieberger,
  ``Explicit BCJ Numerators from Pure Spinors,''
  JHEP {\bf 1107} (2011) 092
  [arXiv:1104.5224 [hep-th]].



\bibitem{Feng:2010my}
  B.~Feng, R.~Huang and Y.~Jia,
  ``Gauge Amplitude Identities by On-shell Recursion Relation in S-matrix Program,''
  Phys.\ Lett.\ B {\bf 695} (2011) 350
  [arXiv:1004.3417 [hep-th]].
\bibitem{Jia:2010nz}
  Y.~Jia, R.~Huang and C.~-Y.~Liu,
  ``$U(1)$-decoupling, KK and BCJ relations in $\mathcal{N}=4$ SYM,''
  Phys.\ Rev.\ D {\bf 82}  (2010) 065001
  [arXiv:1005.1821 [hep-th]].
\bibitem{Chen:2011jxa}
  Y.~-X.~Chen, Y.~-J.~Du and B.~Feng,
  ``A Proof of the Explicit Minimal-basis Expansion of Tree Amplitudes in Gauge Field Theory,''  JHEP {\bf 1102} (2011) 112  [arXiv:1101.0009 [hep-th]].  



  \bibitem{KLT} H. Kawai, D. Lewellen and H. Tye, "A Relation Betwwen Tree
Amplitudes of Closed and Open Strings", Nucl.Phys.B269 (1986)1.


\bibitem{Bern:1998ug}
  Z.~Bern, L.~J.~Dixon, D.~C.~Dunbar, M.~Perelstein and J.~S.~Rozowsky,
  ``On the relationship between Yang-Mills theory and gravity
  and its implication for ultraviolet divergences,''
  Nucl.\ Phys.\ B {\bf 530} (1998) 401
  [hep-th/9802162].

\bibitem{Bern:2010yg}
  Z.~Bern, T.~Dennen, Y.~-t.~Huang and M.~Kiermaier,
  ``Gravity as the Square of Gauge Theory,''
  Phys.\ Rev.\ D {\bf 82}  (2010) 065003
  [arXiv:1004.0693 [hep-th]].


\bibitem{Bern:2008pv}
  Z.~Bern, J.~J.~M.~Carrasco, L.~J.~Dixon, H.~Johansson and R.~Roiban,
  Phys.\ Rev.\ D {\bf 78}, 105019 (2008)
  [arXiv:0808.4112 [hep-th]].

\bibitem{Bern:2009kd}
  Z.~Bern, J.~J.~Carrasco, L.~J.~Dixon, H.~Johansson and R.~Roiban,
  Phys.\ Rev.\ Lett.\  {\bf 103}, 081301 (2009)
  [arXiv:0905.2326 [hep-th]].

\bibitem{Bern:2010ue}
  Z.~Bern, J.~J.~M.~Carrasco and H.~Johansson,
  ``Perturbative Quantum Gravity as a Double Copy of Gauge Theory,''
  Phys.\ Rev.\ Lett.\  {\bf 105}  (2010) 061602
  [arXiv:1004.0476 [hep-th]].


\bibitem{Carrasco:2011mn}
  J.~J.~.Carrasco and H.~Johansson,
  Phys.\ Rev.\ D {\bf 85}, 025006 (2012)
  [arXiv:1106.4711 [hep-th]].

\bibitem{Bern:2011rj}
  Z.~Bern, C.~Boucher-Veronneau and H.~Johansson,
  Phys.\ Rev.\ D {\bf 84}, 105035 (2011)
  [arXiv:1107.1935 [hep-th]].

\bibitem{BoucherVeronneau:2011qv}
  C.~Boucher-Veronneau and L.~J.~Dixon,
  JHEP {\bf 1112}, 046 (2011)
  [arXiv:1110.1132 [hep-th]].

\bibitem{Bern:2012uf}
  Z.~Bern, J.~J.~M.~Carrasco, L.~J.~Dixon, H.~Johansson and R.~Roiban,
  Phys.\ Rev.\ D {\bf 85}, 105014 (2012)
  [arXiv:1201.5366 [hep-th]].

\bibitem{Bern:2012cd}
  Z.~Bern, S.~Davies, T.~Dennen and Y.~-t.~Huang,
  Phys.\ Rev.\ Lett.\  {\bf 108}, 201301 (2012)
  [arXiv:1202.3423 [hep-th]].

\bibitem{Bern:2012gh}
  Z.~Bern, S.~Davies, T.~Dennen and Y.~-t.~Huang,
  Phys.\ Rev.\ D {\bf 86}, 105014 (2012)
  [arXiv:1209.2472 [hep-th]].

\bibitem{Oxburgh:2012zr}
  S.~Oxburgh and C.~D.~White,
  ``BCJ duality and the double copy in the soft limit,''
  arXiv:1210.1110 [hep-th].


\bibitem{Yuan:2012rg}
  E.~Y.~Yuan,
  arXiv:1210.1816 [hep-th].

\bibitem{Saotome:2012vy}
  R.~Saotome and R.~Akhoury,
  arXiv:1210.8111 [hep-th].

\bibitem{Boels:2012sy}
  R.~H.~Boels and R.~S.~Isermann,
  arXiv:1212.3473 [hep-th].



\bibitem{Monteiro:2011pc}
  R.~Monteiro and D.~O'Connell,
  ``The Kinematic Algebra From the Self-Dual Sector,''
  JHEP {\bf 1107} (2011) 007
  [arXiv:1105.2565 [hep-th]].


\bibitem{BjerrumBohr:2012mg}
  N.~E.~J.~Bjerrum-Bohr, P.~H.~Damgaard, R.~Monteiro and D.~O'Connell,
  ``Algebras for Amplitudes,''
  JHEP {\bf 1206} (2012) 061
  [arXiv:1203.0944 [hep-th]].
















\bibitem{DelDuca:1999rs}
  V.~Del Duca, L.~J.~Dixon and F.~Maltoni,
  ``New color decompositions for gauge amplitudes at tree and loop level,''
  Nucl.\ Phys.\ B {\bf 571} (2000) 51
  [hep-ph/9910563].

\bibitem{Kleiss:1988ne}
  R.~Kleiss and H.~Kuijf,
  ``MULTI - GLUON CROSS-SECTIONS AND FIVE JET PRODUCTION AT HADRON COLLIDERS,''
  Nucl.\ Phys.\  B {\bf 312} (1989) 616.

\bibitem{Du:2011js}
  Y.~-J.~Du, B.~Feng and C.~-H.~Fu,
  ``BCJ Relation of Color Scalar Theory and KLT Relation of Gauge Theory,''
  JHEP {\bf 1108} (2011) 129
  [arXiv:1105.3503 [hep-th]].


\bibitem{Bern:1999bx}
  Z.~Bern, A.~De Freitas and H.~L.~Wong,
  ``On the coupling of gravitons to matter,''
  Phys.\ Rev.\ Lett.\  {\bf 84} (2000) 3531
  [arXiv:hep-th/9912033].

\bibitem{Bern:2011ia}
  Z.~Bern and T.~Dennen,
  ``A Color Dual Form for Gauge-Theory Amplitudes,''  Phys.\ Rev.\ Lett.\
   {\bf 107}, 081601 (2011)  [arXiv:1103.0312 [hep-th]].  







\bibitem{Berends:1987me}
  F.~A.~Berends and W.~T.~Giele,
  ``Recursive Calculations for Processes with n Gluons,''
  Nucl.\ Phys.\ B {\bf 306} (1988) 759.



\bibitem{Dixon:1996wi}
  L.~J.~Dixon,
  ``Calculating scattering amplitudes efficiently,''
  In *Boulder 1995, QCD and beyond* 539-582
  [hep-ph/9601359].





\bibitem{BjerrumBohr:2010ta}
  N.~E.~J.~Bjerrum-Bohr, P.~H.~Damgaard, B.~Feng and T.~Sondergaard,
  ``Gravity and Yang-Mills Amplitude Relations,''
  Phys.\ Rev.\  D {\bf 82}, 107702 (2010)
  [arXiv:1005.4367 [hep-th]].



\bibitem{Berends:1988zn}
  F.~A.~Berends and W.~T.~Giele,
  ``Multiple Soft Gluon Radiation in Parton Processes,''
  Nucl.\ Phys.\ B {\bf 313}, 595 (1989).  


\end{thebibliography}
\end{document}